\documentclass[lineno]{taira}
\usepackage{graphicx}
\usepackage{newtxtext}
\usepackage{newtxmath}
\usepackage{natbib}
\usepackage{hyperref}
\hypersetup{
	colorlinks = true,
	urlcolor   = blue,
	citecolor  = black,
}

\newcommand{\RomanNumeralCaps}[1]
\linenumbers

\usepackage{graphicx,amssymb,amsmath,color,epsfig,soul,rotating,overpic,varwidth,xcolor}

\usepackage[export]{adjustbox}
\usepackage{subfigure}
\usepackage{subfigmat}
\usepackage{algorithm}
\usepackage[noend]{algpseudocode}
\usepackage{bm}
\usepackage{xfrac}
\usepackage{multicol}
\usepackage{multirow}
\usepackage{pbox}
\usepackage{tabularx}

\usepackage{caption}

\usepackage{booktabs}
\usepackage{tikz}
\usetikzlibrary{tikzmark}

\usepackage{lipsum}

\definecolor{blue2}{rgb}{0, 0.4470, 0.7410}
\definecolor{red2}{rgb}{0.8500, 0.1250, 0.0480} 
\definecolor{orange2}{rgb}{0.8500, 0.3250, 0.0980} 
\definecolor{yellow2}{rgb}{0.9290, 0.6940, 0.1250}
\definecolor{purple2}{rgb}{0.4940, 0.1840, 0.5560}
\definecolor{green2}{rgb}{0.4660, 0.6740, 0.1880}
\definecolor{ltblue2}{rgb}{0.3010, 0.7450, 0.9330}
\definecolor{dkred2}{rgb}{0.6350, 0.0780, 0.1840}
\definecolor{gray2}{rgb}{0.22, 0.22, 0.3}
\definecolor{ltgray2}{rgb}{0.647, 0.647, 0.647}
\definecolor{blueIV}{rgb}{0, 0, 0.7410}
\definecolor{blueIII}{rgb}{0.2, 0.2, 0.7410}
\definecolor{blueII}{rgb}{0.4, 0.4, 0.7410}
\definecolor{blueI}{rgb}{0.7410, 0.7410, 0.7410}
\definecolor{jetVI}{rgb}{0.9763    0.9831    0.0538}
\definecolor{jetV}{rgb}{0.9264    0.7256    0.2996}
\definecolor{jetIV}{rgb}{0.4783    0.7489    0.4877}
\definecolor{jetIII}{rgb}{0.0282    0.6663    0.7574}
\definecolor{jetII}{rgb}{0.0582    0.4677    0.8589}
\definecolor{jetI}{rgb}{0.2081    0.1663    0.5292}
\definecolor{mode1}{rgb}{0 0.4470 0.7410}
\definecolor{mode2}{rgb}{0.8500 0.3250 0.0980}
\definecolor{mode3}{rgb}{0.9290 0.6940 0.1250}
\definecolor{dkgold}{rgb}{0.5930 0.5150 0.3260}
\definecolor{matblue}{rgb}{0.000 0.447 0.741}
\definecolor{matred}{rgb}{0.850 0.325 0.098}
\definecolor{matyellow}{rgb}{0.9290 0.6940 0.125}
\definecolor{matpurple}{rgb}{0.494 0.184 0.556}
\definecolor{matgreen}{rgb}{0.466 0.674 0.188}

\newcommand{\bs}{\boldsymbol}

\begin{document}
	
	\captionsetup{font=scriptsize,labelfont=scriptsize}
	
	\shorttitle{Triglobal resolvent analysis of swept-wing wakes} 
	\shortauthor{J. H. M. Ribeiro et al.} 
	
	\title{Triglobal resolvent analysis of swept-wing wakes} 
	
	\author
	{
		Jean H\'{e}lder Marques Ribeiro\aff{1}
		\corresp{\email{jeanmarques@g.ucla.edu}},
		Chi-An Yeh\aff{1,2},
		\and
		Kunihiko Taira\aff{1}
	}
	
	\affiliation
	{
		\aff{1}
		Department of Mechanical and Aerospace Engineering, University of California, Los Angeles, CA 90095, USA
		
		\aff{2}
		Department of Mechanical and Aerospace Engineering, North Carolina State University, Raleigh, NC 27695, USA
		
	}
	
	\maketitle
	
	\begin{abstract}
		Through triglobal resolvent analysis, we reveal the effects of wing tip and sweep angle on laminar separated wakes over swept wings. For the present study, we consider wings with semi-aspect ratios from $1$ to $4$, sweep angles from $0^\circ$ to $45^\circ$, and angles of attack of $20^\circ$ and $30^\circ$ at a chord-based Reynolds number of $400$ and a Mach number of $0.1$.  Using direct numerical simulations, we observe that unswept wings develop vortex shedding near the wing root with a quasi-steady tip vortex. For swept wings, vortex shedding is seen near the wing tip for low sweep angles, while the wakes are steady for wings with high sweep angles. To gain further insights into the mechanisms of flow unsteadiness, triglobal resolvent analysis is used to identify the optimal spatial input-output mode pairs and the associated gains over a range of frequencies. The three-dimensional forcing and response modes reveal that harmonic fluctuations are directed towards the root for unswept wings and towards the wing tip for swept wings. The overlapping region of the forcing-response mode pairs uncovers triglobal resolvent wavemakers associated with self-sustained unsteady wakes of swept wings.  Furthermore, we show that for low aspect ratio wings optimal perturbations develop globally over the entire wingspan. The present study uncovers physical insights on the effects of tip and sweep on the growth of optimal harmonic perturbations and the wake dynamics of separated flows over swept wings.
	\end{abstract}
	
	\section{Introduction}
	\label{sec:intro}
	Understanding flow separation over finite swept wings is essential to the study of aircraft and biological flight \citep{Anderson:10,Videler:Science04,Lentink:Nature07}. The aspect ratio, angle of attack, and sweep play important roles in influencing stall and wake characteristics \citep{Zhang:JFM20b,Zhang:JFM20}. Although a number of studies have deepened our knowledge of laminar separated wakes around swept wings, coherent flow structures associated with the three-dimensional ($3$-D) flow separation have not been characterized in a comprehensive manner. Such findings would be crucial to explain the role played by the perturbations in characterizing the wakes and support efforts to control flow separation around finite wings.
	
	Previous studies have shown the effect of sweep on post-stall wake characteristics with focus on the role of spanwise flow over wings \citep{Harper:64}. The spanwise flow induced by sweep delays the emergence of stall \citep{Yen:AIAAJ07,Yen:JFE09} and reduces wake oscillations, as shown for high-Reynolds number flows over transonic buffets in biglobal \citep{Crouch:JFM19,Paladini:PRF19,Plante:JFM21} and triglobal linear stability analysis \citep{Timme:JFM20,HeTimme:JFM21}.  Similar observations have been made for flows around aircraft models in experiments \citep{Masini:JFM20} and computations \citep{Houtman:AIAA22}.
	
	At a low Reynolds number, direct numerical simulations (DNS) from \cite{Zhang:JFM20b} showed that sweep angle can significantly alter the wake patterns. For wings with low sweep angles, vortex shedding develops near the wing tip, while unsteadiness is suppressed for flows over highly swept wings. Similar attenuation of flow unsteadiness was further studied for forward-swept wings \citep{Zhang:PRF22} revealing that wing sweep has a strong effect on attenuating wake oscillations in laminar flows. Furthermore, linear instabilities around swept wings were examined for a variety of swept and unswept wings, showing that the sweep angle suppresses the emergence of wake modes \citep{Burtsev:JFM22,Ribeiro:AIAA22}.
	
	The aspect ratio of the wing also affects the wake dynamics on separated flows due to the wing tip vortex in steady \citep{Devenport:JFM96,Torres:AIAAJ04,Taira:JFM09} and unsteady wing motion \citep{Buchholz:JFM06,Yilmaz:JFM12}.  For low-aspect-ratio wings, the tip vortex may suppress the leading-edge vortex formation, reducing the wake unsteadiness \citep{Taira:JFM09}. Tip vortices can also produce adverse effects on the wing, with induced drag and a reduced lift.
	
	To alter the wake dynamics with a proper actuation input, we need to identify the optimal forcing structures that can be amplified  in the flow field \citep{Edstrand:JFM18a,Edstrand:JFM18b}. For this task, we may use modal analysis techniques \citep{Taira:AIAAJ17,Taira:AIAAJ20} to study the dynamics of flow oscillations. Resolvent analysis is an attractive tool for the present study because it identifies the optimal input perturbations in the flow field, their energy amplification, and the characteristics of their unsteady response \citep{Trefethen:93,Jovanovic:JFM05}. Furthermore, with the diverse steady and unsteady wakes observed around swept wings, resolvent analysis can provide a comprehensive study of the input-output dynamics around wings with different aspect ratios, angles of attack, and sweep.
	
	Resolvent analysis has been used to study a broad range of fluid flows \citep{Moarref:JFM13,Thomareis:PRF18,Schmidt:JFM18,Skene:JFM19,Yeh:PRF20,Ricciardi:JFM22}. This approach was initially formulated for steady base flows, to identify modal structures that can be amplified in stable flow regimes \citep{Trefethen:93}. This perspective on fluid dynamics was later extended to unstable systems by \cite{Jovanovic:JFM05} and to unsteady and turbulent flows by \cite{McKeon:JFM10}. In these formulations, a time-averaged flow is used as a base state and nonlinear terms act as sustained forcing in the flow field. In both steady and unsteady flows, resolvent analysis identifies harmonic forcings that produce an amplified response in the flow.
	
	In this study, we identify the optimal spatial input-output modes around the wing through a $3$-D global (triglobal) resolvent analysis, that assumes no spatial homogeneity. Moreover, we gain insights into the self-sustained fluctuations that support unsteadiness on laminar separated flows using resolvent wavemakers, which are similar in spirit to the eigenvector-based wavemakers \citep{Giannetti:JFM07,Giannetti:JFM10}. The resolvent wavemakers, also named as structural sensitivity, are obtained from the overlap of forcing and response modes \citep{Qadri:PRF17,Skene:JFM22}. These findings provide a comprehensive analysis of the energy amplification mechanisms in flows around swept wings through an input-output process, identifying the optimal locations where perturbations can be introduced to alter the wake behavior. Therefore, these findings are crucial for the development of efficient flow control strategies  \citep{Yeh:JFM19,Liu:JFM21} that aim to improve the aerodynamic performance of swept wings experiencing massive flow separation.
	
	The present paper on triglobal resolvent analysis is organized as follows. In section \ref{sec:problem}, we describe the problem setup for the current work. In section \ref{sec:results}, we discuss our main findings from triglobal resolvent analysis. We identify the emergence of wake unsteadiness caused by the overlap of optimal forcing and response modes in the near wake. Perturbations are directed towards the region where vortex shedding takes place. The locations of the optimal forcing and response modes over the wingspan also suggest that wakes of highly swept wings are more resilient to external perturbations. Furthermore, we find that low-aspect-ratio wings limit the growth of perturbations to global modes extending over the entire wingspan. Finally, our conclusions are presented in section \ref{sec:conclusions}.

	\section{Problem setup}
	\label{sec:problem}
	
	\begin{figure}
		\footnotesize
		\centering
		\begin{tikzpicture}
		\node[anchor=south west,inner sep=0] (image) at (0.0,0.0) {\includegraphics[trim=0mm 2mm 2mm 0mm, clip,width=1.0\textwidth]{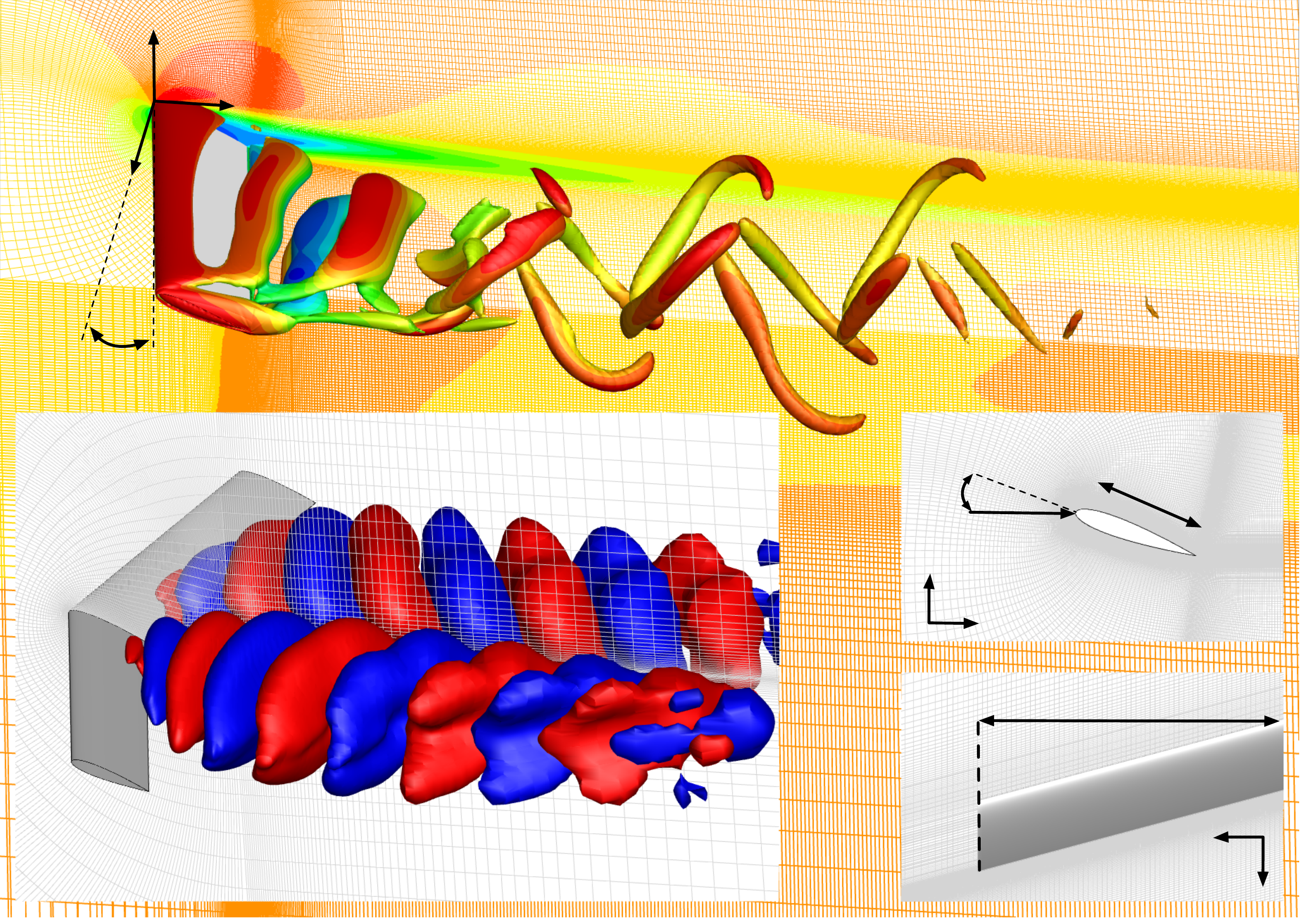}};
		\node[align=left] at (1.20,5.70) {\rotatebox{0}{$\Lambda$}};
		\node[align=left] at (2.50,8.60) {\rotatebox{0}{$x$}};
		\node[align=left] at (1.80,9.10) {\rotatebox{0}{$y$}};
		\node[align=left] at (1.20,8.00) {\rotatebox{0}{$z$}};
		\node[align=left] at (9.85,3.70) {\rotatebox{0}{$y$}};
		\node[align=left] at (10.35,3.15) {\rotatebox{0}{$x$}};
		\node[align=left] at (10.70,3.85) {\rotatebox{0}{$U_\infty$}};
		\node[align=left] at (12.10,4.55) {\rotatebox{0}{$c$}};
		\node[align=left] at (9.80,4.40) {\rotatebox{0}{$\alpha$}};
		\node[align=left] at (12.95,0.35) {\rotatebox{0}{$x$}};
		\node[align=left] at (12.45,0.85) {\rotatebox{0}{$z$}};
		\node[align=left] at (12.00,2.20) {\rotatebox{0}{$b$}};
		\scriptsize
		\draw[gray, very thick] (0.0,0.0) rectangle (13.50,9.55);
		\draw[rounded corners,gray, fill=white, very thick] (1.0,9.35) rectangle (3.0,9.75);
		\node[align=left] at (2.0,9.55) {DNS domain};
		\draw[gray, very thick] (9.40,2.85) rectangle (13.40,5.25);
		\draw[rounded corners,gray, fill=white, very thick] (9.60,5.05) rectangle (11.30,5.45);
		\node[align=left] at (10.45,5.22) {($x$,$y$) plane};
		\draw[gray, very thick] (9.40,0.15) rectangle (13.40,2.50);
		\draw[rounded corners,gray, fill=white, very thick] (9.60,2.30) rectangle (11.30,2.70);
		\node[align=left] at (10.45,2.48) {($x$,$z$) plane};
		\draw[blue2, very thick] (0.20,0.15) rectangle (8.10,5.25);
		\draw[rounded corners,blue2, fill=white, very thick] (1.0,5.05) rectangle (3.00,5.45);
		\node[align=left] at (2.05,5.22) {resolvent analysis};
		\node[anchor=south west,inner sep=0] (image) at (12.2,8.2) {\includegraphics[frame,trim=0mm 0mm 0mm 0mm, clip,width=0.015\textwidth,height=0.05\textwidth]{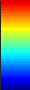}};
		\tiny
		\node[align=left] at (12.7,8.25) {\rotatebox{0}{$-0.2$}};
		\node[align=left] at (12.6,8.85) {\rotatebox{0}{$1.0$}};
		\node[align=left] at (12.35,9.05) {\rotatebox{0}{$u_x$}};
		\end{tikzpicture} 
		\caption{Setup for finite swept wing simulation. In the gray boxes, the instantaneous flow field for $\alpha = 20^\circ$, $\Lambda = 15^\circ$, and $sAR = b/c = 4$, with $Q = 2$ isosurfaces colored by instantaneous $u_x$. Mesh colored by time-averaged $\bar{u}_x$. In the blue box, isosurfaces of the primary response mode with mesh in light gray.} 
		\label{fig:setup_mesh}
	\end{figure}
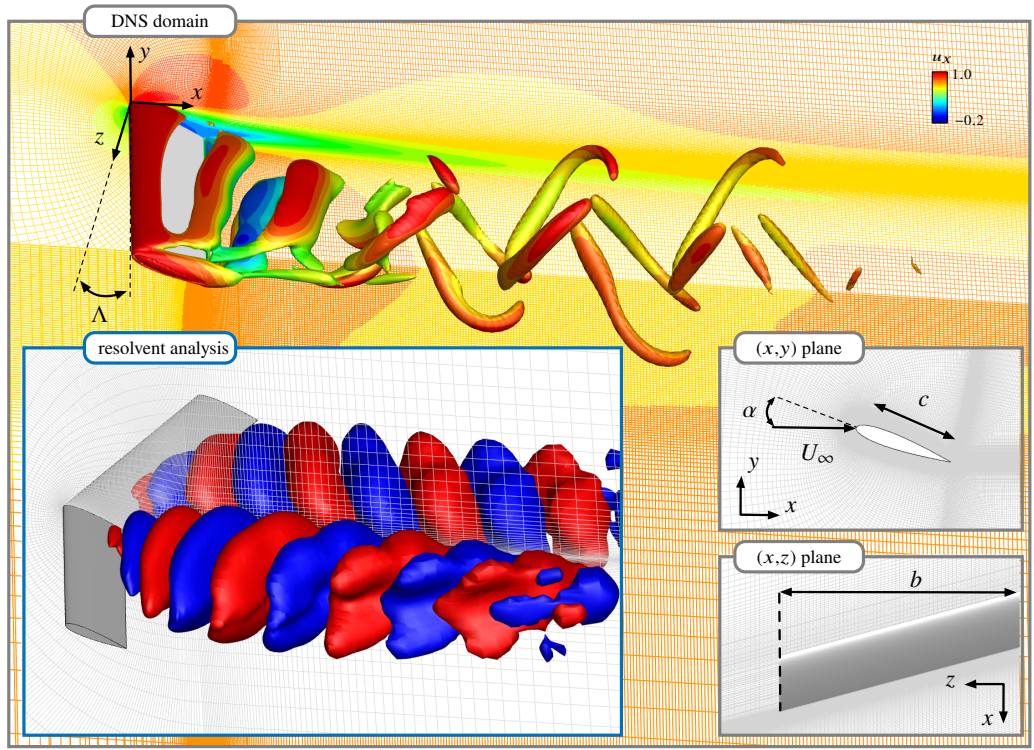
	
	We consider laminar flows over untapered swept wings with NACA 0015 cross-sectional profile, as shown in figure \ref{fig:setup_mesh}. The spatial coordinates are defined with $(x,y,z)$ being the streamwise, transverse, and spanwise directions, respectively, with the origin placed at the leading edge of the wing root. The NACA 0015 airfoil geometry is defined on the $(x,y)$ plane. The wingspan is formed by extruding the airfoil profile in the spanwise direction. The semi-aspect ratio is defined through the half-span length $b$ and the chord-length $c$ as $sAR = b/c$, with values set between $1 \le sAR \le 4$. For swept wings, the $3$-D computational setup is sheared in the $x$-direction and the sweep angle is defined between the $z$-direction  and the leading edge. In the present work, we consider sweep angles $0 \le \Lambda \le 45^\circ$. The angle of attack, $\alpha = 20^\circ$ and $30^\circ$, is defined between the streamwise direction and the airfoil chord line. To focus on the effects of wing tip and sweep in the wake dynamics, we analyze a half-span model with symmetry boundary conditions imposed at the root plane. The wings have a straight-cut tip and sharp trailing edge.  For all flows analyzed herein, we define the chord-based Reynolds number $Re_{c} = U_\infty c/\nu = 400$, where $U_\infty$ is the freestream velocity and $\nu$ is the kinematic viscosity. The freestream Mach number is set to $M_\infty   = U_\infty / a_\infty  = 0.1$, where $a_\infty$ is the freestream speed of sound.
	
	\subsection{Direct numerical simulation}
	\label{sec:problem_dns}
	
	We perform DNS with the compressible flow solver \textit{CharLES} \citep{Khalighi:AIAA11,Bres:AIAAJ17}, which uses a second-order accurate finite-volume method in space with a third-order accurate scheme in time. With the origin at the leading edge of the airfoil $(x/c,\ y/c,\ z/c) = (0,0,0)$, the computational domain extends over $(x/c, y/c, z/c) \in [-20,25] \times  [-20,20] \times [0,20]$. We build a C-type grid for each angle of attack with $\min$ $(\Delta x, \Delta y, \Delta z)/c = (0.005, 0.005, 0.04)$ applying mesh refinement near the airfoil and in the wake, as shown in~figure~\ref{fig:setup_mesh}.

	Inlet and farfield boundaries are prescribed with Dirichlet boundary conditions $(\rho, u_x, u_y, u_z, p) = (\rho_\infty, U_\infty, 0, 0, p_\infty)$, where $\rho$ is density, $p$ is pressure, and $u_x$, $u_y$, and $u_z$ are velocity components in the $x$, $y$, and $z$ directions, respectively. Variables with subscript $\infty$ denote freestream values. For all setups considered herein, the velocity boundary conditions applied at the inlet and farfield are aligned with the $x$-direction, which enforces the same streamwise flow over wings with different sweep angles. The airfoil surface is provided with adiabatic no-slip boundary condition. To simulate a half-wing model, we prescribe the symmetry boundary condition along the root plane. A sponge layer is applied at the outlet over $x/c \in [15,25]$ with the target state being the running-averaged flow over $5$ convective time units $t \equiv c / U_\infty$ \citep{Freund:AIAAJ97}. Simulations start with uniform flow and time integration is performed with a constant acoustic Courant--Friedrichs--Lewy (CFL) number of $1$. After transients are flushed out of the computational domain, the time-averaged base flow $\bar{\mathbi{q}}$ is determined over $50$ convective time units. The present results were carefully verified and validated. Close agreement for instantaneous and time-averaged velocity components was achieved with those from \cite{Zhang:JFM20b}. We have further validated our computations for time-averaged drag and lift coefficients,
	\begin{equation}
		C_D =  \frac{F_x}{\frac{1}{2} \rho U_\infty^2 b c} \quad \text{and} \quad C_L =  \frac{F_y}{\frac{1}{2} \rho U_\infty^2 b c}  \mbox{   ,}
	\end{equation}
	respectively, where $F_x$ is the drag and $F_y$ is the lift over the wing, as reported in table \ref{tab:validation}.
	
	\renewcommand{\arraystretch}{1.2}
	\begin{table}
		\vspace*{-3mm}
		\centering
		\begin{tabular}{p{1.15in}p{0.37in}p{0.37in}p{0.001in}p{0.37in}p{0.37in}p{0.001in}p{0.37in}p{0.37in}p{0.001in}p{0.37in}p{0.37in}}
			\multicolumn{1}{r}{} &
			\multicolumn{2}{c}{\hspace{-2mm}$\Lambda = 0^\circ$} &&
			\multicolumn{2}{c}{\hspace{-2mm}$\Lambda = 15^\circ$} &&
			\multicolumn{2}{c}{\hspace{-2mm}$\Lambda = 30^\circ$} &&
			\multicolumn{2}{c}{\hspace{-2mm}$\Lambda = 45^\circ$} \\
			& 
			\multicolumn{1}{l}{\hspace{0mm}$\overline{C_L}$} & \multicolumn{1}{l}{\hspace{0mm}$\overline{C_D}$} && 
			\multicolumn{1}{l}{\hspace{0mm}$\overline{C_L}$} & \multicolumn{1}{l}{\hspace{0mm}$\overline{C_D}$} &&
			\multicolumn{1}{l}{\hspace{0mm}$\overline{C_L}$} & \multicolumn{1}{l}{\hspace{0mm}$\overline{C_D}$} &&
			\multicolumn{1}{l}{\hspace{0mm}$\overline{C_L}$} & \multicolumn{1}{l}{\hspace{0mm}$\overline{C_D}$} \\ 
			Present study & $0.53$ & $0.35$ && $0.50$ & $0.34$ && $0.45$ & $0.31$ && $0.40$ & $0.29$ \\
			\cite{Zhang:JFM20b} & $0.53$ & $0.35$ && $0.51$ & $0.33$ && $0.44$ & $0.30$ && $0.40$ & $0.29$  \\
		\end{tabular} \vspace*{-2mm}
		\caption{Time-averaged lift and drag coefficients ($\overline{C_L}$ and $\overline{C_D}$) compared to \cite{Zhang:JFM20b} for laminar separated flow over NACA 0015 wings with $sAR = 4$, $\alpha = 20^\circ$, and $\Lambda = 0^\circ$, $15^\circ$, $30^\circ$, and $45^\circ$. }
		\label{tab:validation}
	\end{table}	
	
	\begin{figure}
		\begin{tikzpicture}
		\node[anchor= west,inner sep=0] (image) at (0.0,0.0) {\includegraphics[trim=0mm 0mm 0mm 0mm, clip,width=1.00\textwidth]{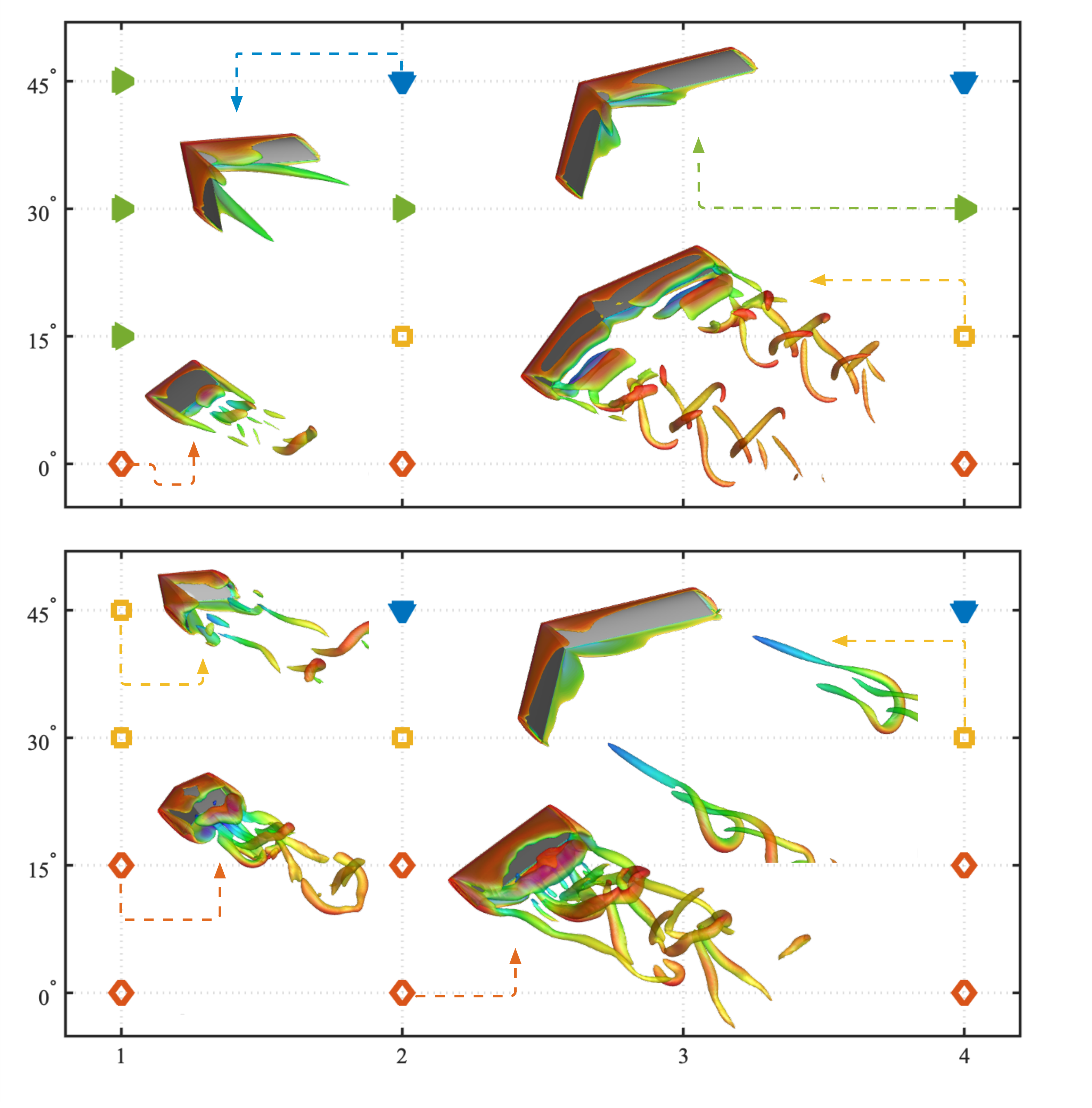}};
		\scriptsize
		\node[align=left] at (0.0,3.5) {\rotatebox{90}{$\Lambda$}};
		\node[align=left] at (0.0,-3.0) {\rotatebox{90}{$\Lambda$}};
		\node[align=left] at (1.20,6.80) {\rotatebox{0}{$\alpha = 20^\circ$}};
		\node[align=left] at (1.20,0.20) {\rotatebox{0}{$\alpha = 30^\circ$}};
		\node[align=left] at (7.00,-6.7) {\rotatebox{0}{$sAR$}};
		\node[anchor=south west,inner sep=0] (image) at (12.9,-0.50) {\includegraphics[frame,angle=0,trim=0mm 0mm 0mm 0mm, clip,width=0.017\textwidth,height=0.08\textwidth]{figs/JFM_jet_cb.png}};
		\tiny
		\node[align=left] at (13.4,-0.50) {\rotatebox{0}{$-0.2$}};
		\node[align=left] at (13.35,0.55) {\rotatebox{0}{$1.0$}};
		\node[align=left] at (13.05,0.80) {\rotatebox{0}{$u_x$}};
		\end{tikzpicture} \vspace{-5mm}
		\caption{Instanteneous isosurfaces of $Q = 2$ colored by $u_x$. Unsteady shedding near wing root ({\color{matblue} {\large $\bs{\diamond}$}}), unsteady shedding near wing tip  ({\color{matyellow} $\bs{\square}$}), steady flow with root structures ({\color{matgreen} $\bs{\blacktriangleright}$}), steady flow with streamwise vortices ({\color{matblue} $\bs{\blacktriangledown}$}).}
		\label{fig:validation}
	\end{figure}
	
	A variety of wake patterns can be observed for different $\alpha$, $\Lambda$, and $sAR$, as summarized in figure \ref{fig:validation}. In the bottom plot, the flow over the wing with $(sAR,\alpha,\Lambda) = (2,30^\circ,0^\circ)$ exhibits a quasi-steady streamwise oriented tip vortex. This structure is characteristic of flows over unswept wings and also appears around wings with different $\alpha$ and $sAR$. For such wings, unsteady spanwise vortices develop at the root plane. Between the root and the wing tip, there is an intermediate zone with braid-like vortices.
	
	Wing sweep affects the wake dynamics and structures. At low sweep angles, a spanwise flow develops over the wing and advects unsteady vortices towards the wing tip. For instance, for $(sAR,\alpha,\Lambda) = (4,20^\circ,15^\circ)$, spanwise vortices still appear. These structures are similar to the ones observed over unswept wings, although they form closer to the wing tip, and break into helical structures in the wake. For the flow around this wing, streamwise-oriented tip vortices are absent.
	
	For higher $\Lambda$, wing sweep can attenuate wake oscillations. For $(sAR,\alpha,\Lambda) = (4,30^\circ,30^\circ)$ near-wake unsteadiness is reduced and unsteady vortices appear further downstream in the wake. We notice that these structures are absent when the angle of attack is lowered to $20^\circ$. Similarly, increasing the sweep angle to $\Lambda = 45^\circ$ suppresses unsteady vortices on both angles of attack and the wake becomes steady. On such highly swept wings, ram-horn shaped streamwise-oriented vortices develop from the root plane and extend into the wake.  
	
	For each $(\alpha,\Lambda)$ pair, the wake exhibits similar characteristics for wings with $sAR = 4$ and $2$. Reducing the semi-aspect ratio to $sAR = 1$ has a strong influence on the wake dynamics, as shown in figure \ref{fig:validation}. For such wings, tip effects can suppress the formation of leading edge vortices at lower angles of attack. For instance, at $\alpha = 20^\circ$, wake unsteadiness is reduced and swept wings exhibit steady flows with root structures. 
	
	Unsteady vortices are observed in flows over $sAR = 1$ wings at $\alpha = 30^\circ$ for all considered sweep angles. The unsteadiness appears near the root for lower $\Lambda$. Further downstream, unsteady vortices appear over the entire wingspan. For higher $\Lambda$, vortices are generated near the wing tip and helical structures are observed in the wake. These observations agree with the characterizations by \cite{Zhang:JFM20b}. To deepen our insights into swept-wing wake dynamics, we now call for the triglobal resolvent analysis.
	
	\subsection{Resolvent analysis}
	\label{sec:problem_resolvent}
	
	Let us consider the Reynolds decomposition of state variable $\mathbi{q} = \bar{\mathbi{q}} + \mathbi{q}^\prime$, where $\bar{\mathbi{q}}$ is the time-averaged flow and $\mathbi{q}'$ is the statistically stationary fluctuation component \citep{McKeon:JFM10}. This decomposition along with spatial discretization is used to linearize the compressible Navier--Stokes equations about $\bar{\boldsymbol{q}}$ to yield
	\begin{equation}
	\frac{\partial \mathbi{q}^\prime}{\partial t} = \mathsfbi{L}_{\bar{\mathbi{q}}}\mathbi{q}^\prime + \mathbi{f}^\prime \mbox{   ,}
	\label{eq:LNS}
	\end{equation}
	where $\mathsfbi{L}_{\bar{\mathbi{q}}}$ is the discrete linearized Navier--Stokes operator \citep{Sun:JFM17} and  $\mathbi{f}^\prime$ accounts for the external forcing and nonlinear terms. With the Fourier representation 
	\begin{equation}
	\left[ \mathbi{q}^\prime(\mathbi{x},t), \mathbi{f}^\prime(\mathbi{x},t) \right] = \int_{-\infty}^{\infty} \left[ \hat{\mathbi{q}}_{\omega}(\mathbi{x}), \hat{\mathbi{f}}_{\omega}(\mathbi{x}) \right] e^{-i \omega t } {\rm d} \omega \mbox{   ,}
	\label{eq:fourier}
	\end{equation}
	we obtain
	\begin{equation}
	-i \omega \hat{\mathbi{q}}_{\omega} = \mathsfbi{L}_{\bar{\mathbi{q}}}\hat{\mathbi{q}}_{\omega} + \hat{\mathbi{f}}_{\omega} \mbox{   ,}
	\label{eq:LNS_fourier}
	\end{equation}
	where $\mathbi{x} = (x,y,z)$ and the triglobal response and forcing modes are $\hat{\mathbi{q}}_{\omega}$ and $\hat{\mathbi{f}}_{\omega}$, respectively, for a temporal frequency $\omega$. This expression leads to 
	\begin{equation}
	\hat{\mathbi{q}}_{\omega} = \mathsfbi{H}_{\bar{\mathbi{q}},\omega} \hat{\mathbi{f}}_{\omega} \mbox{   ,}
	\label{eq:cont_resolvent_1}
	\end{equation}
	in which the resolvent operator $\mathsfbi{H}_{\bar{\mathbi{q}},\omega} \in \mathbb{C}^{m \times m}$, with $m$ defined by the product of the number of state variables and the number of spatial grid points. For the present triglobal base flows, the linear operators have size $m$ between $3$ and $5 \times 10^6$. We analyze the resolvent operator with the singular value decomposition (SVD)
	\begin{equation}
	\mathsfbi{H}_{\bar{\mathbi{q}}} = \left[ -i \omega \mathsfbi{I}  - \mathsfbi{L}_{\bar{\mathbi{q}}} \right]^{-1} = \mathsfbi{Q} \mathbi{\Sigma} \mathsfbi{F}^*,
	\end{equation}
	where $\mathsfbi{F} = [\hat{\mathbi{f}}_1,\hat{\mathbi{f}}_2,\dots,\hat{\mathbi{f}}_m]$ is an orthonormal matrix holding the forcing modes, $\bs{\Sigma} = {\rm diag}[\sigma_1,\sigma_2,\dots,\sigma_m]$ is the diagonal matrix with singular values (gain) in descending order, and $\mathsfbi{Q} =  [\hat{\mathbi{q}}_1,\hat{\mathbi{q}}_2,\dots,\hat{\mathbi{q}}_m]$ is the orthonormal matrix comprised of the response modes \citep{Trefethen:93,Jovanovic:JFM05}. For visualization purposes, we show only the real part of the complex-valued resolvent modes. Here, we employ the Chu norm \citep{Chu:Acta65} incorporating it within the resolvent operator through a similarity transform $\mathsfbi{H}_{\bar{\mathbi{q}}} \rightarrow \mathsfbi{W}^{\frac{1}{2}} \mathsfbi{H}_{\bar{\mathbi{q}}} \mathsfbi{W}^{-\frac{1}{2}}$, where $\mathsfbi{W}$ is the weight matrix  that accounts for numerical quadrature and energy weights. 
	
	Resolvent analysis requires careful consideration of the eigenvalues of $\mathsfbi{L}_{\bar{\mathbi{q}}}$. In the presence of unstable modes in the linear operator eigenspectrum, the asymptotic input-output relationship is buried under the unstable dynamics behavior. The present resolvent analysis utilizes a time-averaged flow as the base state. Since such flow is not the equilibrium state, stability characterization cannot be performed in a strict sense. However, it is important to check the location of the eigenvalues in the complex plane to capture the growth rate of the most unstable modes of $\mathsfbi{L}_{\bar{\mathbi{q}}}$.
	
	To use resolvent analysis to study the wake dynamics of unstable base flows, we examine the dynamics through the lens of temporal discounting \citep{Jovanovic:04}.  Discounting applies a temporal damping on forcing and response modes as $[\hat{\mathbi{q}}_{\omega},\hat{\mathbi{f}}_{\omega}]e^{-\beta t}$, where $\beta>0$ is a time-discounting parameter defined within the discounted resolvent operator \citep{Jovanovic:04}. With the discounted resolvent analysis, we can examine amplification dynamics that takes place on a time scale shorter than that of the most unstable mode. Detailed discussions on our choice of $\beta$ are provided in appendix \ref{sec:appendixDiscounting}. Through the discounted resolvent analysis, valuable insights have been provided in past studies for the dynamics and control of flows over airfoils \citep{Yeh:JFM19,Yeh:PRF20,Ricciardi:JFM22,Ribeiro:JFM22}.
	
	The $\mathsfbi{H}_{\bar{\mathbi{q}},\omega}$ operators were discretized over $3$-D structured grids with the leading edge at the root positioned at $(x/c,\ y/c,\ z/c) = (0,0,0)$, extending over $(x/c, y/c, z/c) \in [-10,15] \times  [-10,10] \times [0,10]$ with near wake grids shown at the bottom left of figure \ref{fig:setup_mesh}. The computational grids used for resolvent analysis have a smaller domain size than those used for DNS. For the base flow on the mesh for resolvent analysis, we perform a linear interpolation of the flow field from DNS mesh to the resolvent mesh. We prescribe homogeneous Neumann boundary conditions for $T'$ and homogeneous Dirichlet boundary conditions for the fluctuating variables $\rho'$ and $u'$ along the farfield, airfoil surface, and outlet. Sponges are applied far from the airfoil and in conjunction with the boundary conditions \citep{Freund:AIAAJ97}.
	
	For the large linear operators in the present work, efficient numerical tools are needed for SVD \citep{Halko:SIAM11}.  We use the randomized resolvent analysis algorithm from \cite{Ribeiro:PRF20},  sketching $\mathsfbi{H}_{\bar{\mathbi{q}},\omega}$ with $10$ random test vectors. Each entry of the test vectors is associated with a particular grid point and the five state variables, scaled by $[\| \nabla \rho\|, \|\nabla u_x\|, \|\nabla u_y\|, \|\nabla u_z\|, \|\nabla T\|]$ at each spatial location for each state variable \citep{Ribeiro:PRF20,House:AIAA22}.  A convergence analysis of the randomized resolvent algorithm is provided in appendix \ref{sec:appendixConvergence}.
	
	The computation of resolvent modes for large linear operators can be challenging for the resolvent analysis of high-Reynolds number flows that require a large grid. The bottleneck is related to the time and memory requirements of the linear systems solvers within the SVD. Building an optimal basis to avoid linear system solvers is possible \citep{Barthel:PRF22}, although a generalization for complex geometries is still challenging. It is possible, however, to obtain accurate resolvent modes with time-stepping instead of direct solvers. Those methods tend to penalize the computational time costs, although a considerable reduction in memory requirements can be achieved \citep{Barkley:IJNMF08,Monokrousos:JFM10,Gomez:JFM16}. The computational time required by time-steppers can also be reduced by incorporating streaming discrete Fourier transforms \citep{Martini:JFM21,Farghadan:AIAA21}. The use of iterative solvers has shown promising results to compute resolvent modes around a commercial aircraft model \citep{Houtman:AIAA22}.
	
	In the present work, the direct and adjoint linear systems were directly solved using the MUMPS (multifrontal massively parallel sparse direct solver) package \citep{Amestoy:SIAM01}. Moreover, we incorporate the adjoint-based sensitivity analysis to interpolate the resolvent norm over frequencies $\omega$ \citep{Schmid:AMR14,Fosas:JoT17}. This approach is used to calculate the gradient of $\sigma$ with respect to $\omega$, allowing an accurate interpolation among frequencies \citep{Skene:JFM19}. The codes used to compute the resolvent modes are part of the `linear analysis package' made available by \cite{SkeneRibeiro:tools}.
	
	\section{Triglobal resolvent analysis}
	\label{sec:results}
	\subsection{Forcing and response modes structures}
	\label{sec:results_switch}
	
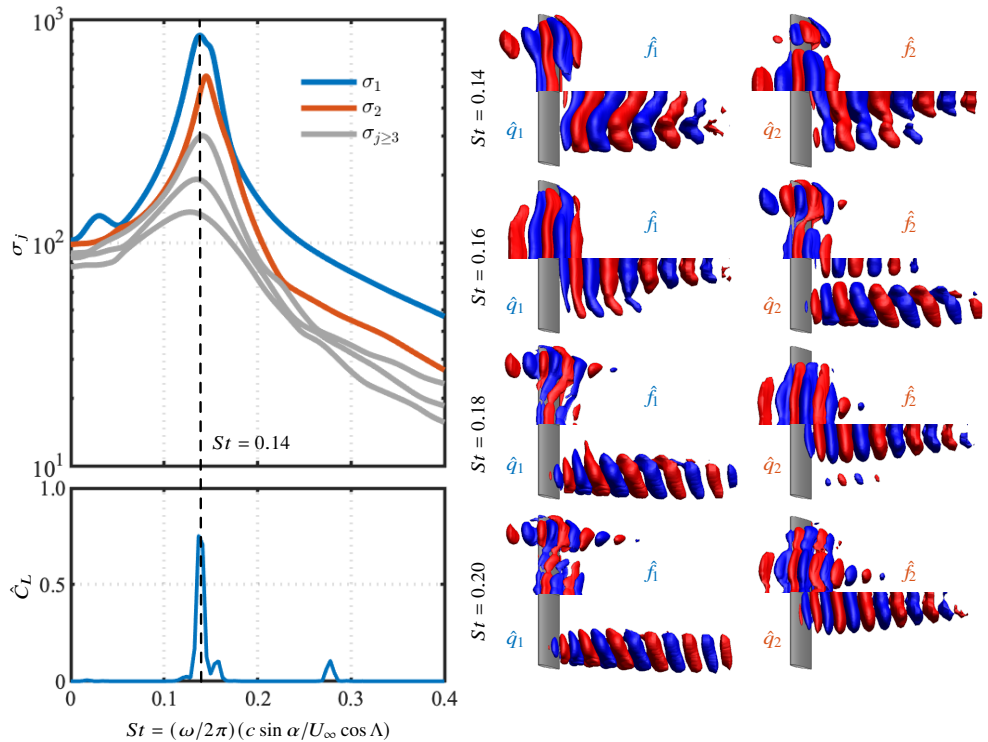
\begin{figure}
		\centering
		\begin{tikzpicture}
		\node[anchor=south west,inner sep=0] (image) at (0,0) {\includegraphics[trim=0mm 0mm 0mm 0mm, clip,width=1.0\textwidth]{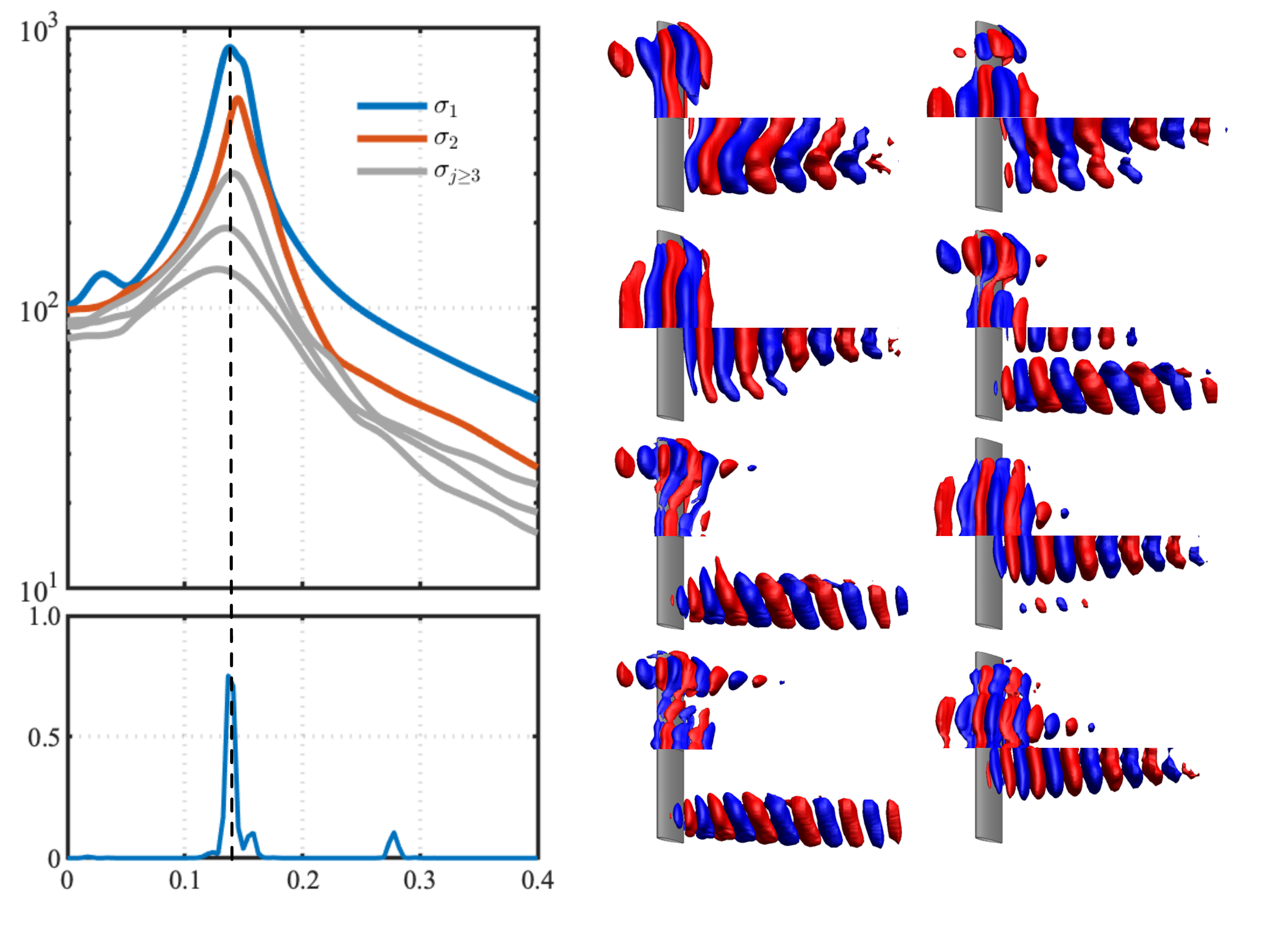}};
		\scriptsize
		\node[align=left] at (0.05,6.6) {\rotatebox{90}{$\sigma_j$}};
		\node[align=left] at (0.05,2.1) {\rotatebox{90}{$\hat{C}_L$}};
		\node[align=left] at (3.2,0.2) {\rotatebox{0}{$St = (\omega / 2\pi) (c \sin \alpha / U_\infty \cos \Lambda)$}};
		\node[align=left] at (3.10,4.0) {\rotatebox{0}{$St = 0.14$}};
		\node[align=left] at (8.4,9.2) {\rotatebox{0}{\color{matblue}$\hat{f}_1$}};
		\node[align=left] at (6.6,8.1) {\rotatebox{0}{\color{matblue}$\hat{q}_1$}};
		\node[align=left] at (8.4,6.9) {\rotatebox{0}{\color{matblue}$\hat{f}_1$}};
		\node[align=left] at (6.6,5.8) {\rotatebox{0}{\color{matblue}$\hat{q}_1$}};
		\node[align=left] at (8.4,4.6) {\rotatebox{0}{\color{matblue}$\hat{f}_1$}};
		\node[align=left] at (6.6,3.7) {\rotatebox{0}{\color{matblue}$\hat{q}_1$}};
		\node[align=left] at (8.4,2.3) {\rotatebox{0}{\color{matblue}$\hat{f}_1$}};
		\node[align=left] at (6.6,1.4) {\rotatebox{0}{\color{matblue}$\hat{q}_1$}};
		\node[align=left] at (11.8,9.2) {\rotatebox{0}{\color{matred}$\hat{f}_2$}};
		\node[align=left] at (10.0,8.1) {\rotatebox{0}{\color{matred}$\hat{q}_2$}};
		\node[align=left] at (11.8,6.9) {\rotatebox{0}{\color{matred}$\hat{f}_2$}};
		\node[align=left] at (10.0,5.8) {\rotatebox{0}{\color{matred}$\hat{q}_2$}};
		\node[align=left] at (11.8,4.6) {\rotatebox{0}{\color{matred}$\hat{f}_2$}};
		\node[align=left] at (10.0,3.7) {\rotatebox{0}{\color{matred}$\hat{q}_2$}};
		\node[align=left] at (11.8,2.3) {\rotatebox{0}{\color{matred}$\hat{f}_2$}};
		\node[align=left] at (10.0,1.4) {\rotatebox{0}{\color{matred}$\hat{q}_2$}};
		\node[align=left] at (6.1,8.4) {\rotatebox{90}{$St = 0.14$}};
		\node[align=left] at (6.1,6.3) {\rotatebox{90}{$St = 0.16$}};
		\node[align=left] at (6.1,4.1) {\rotatebox{90}{$St = 0.18$}};
		\node[align=left] at (6.1,1.9) {\rotatebox{90}{$St = 0.20$}};
		\end{tikzpicture} \vspace{-6mm}
		\caption{Resolvent gains and forcing-response mode pairs for $(sAR,\alpha,\Lambda) = (4,20^\circ,0^\circ)$.  For each mode, forcing ($\hat{\mathbi{f}}$) is the top-half while  response ($\hat{\mathbi{q}}$) it the bottom-half with isosurfaces of velocity $\hat{u}_y \in [-0.2,0.2]$, with freestream directed to the right. On bottom left, power spectrum density of lift coefficient $\hat{C}_L$.} 
		\label{fig:modeSwitching}
	\end{figure}

	Let us first examine the dominant gains, forcing, and response modes for $(sAR,\alpha,\Lambda) = (4,20^\circ,0^\circ)$, as shown in figure \ref{fig:modeSwitching}. The dominant resolvent modes are observed at $St = 0.14$, where 
	\begin{equation}
	St = \frac{\omega}{2\pi} \frac{c \sin \alpha}{ U_\infty \cos \Lambda} \mbox{   }
	\label{eq:strouhal}
	\end{equation}
	is the Fage--Johansen Strouhal number \citep{FageJohanssen:PRSA27} with a $1/\cos \Lambda$ scaling that incorporates the influence of the sweep angle. This frequency scaling is inspired by the independence principle \citep{Wygnanski:JFM11} and collapses the spectral behavior of the resolvent modes over different sweep angles \citep{Ribeiro:JFM22}. This frequency matches the peak frequency for the lift coefficient shown in figure \ref{fig:modeSwitching} (bottom left). The dominant frequency for $\sigma_1$ and $\hat{C}_L$ agrees for all unsteady flows presented herein.
	
	The spatial structures of forcing-response mode pairs are shown in figure \ref{fig:modeSwitching} (right) for representative frequencies. For $St = 0.14$, primary modes exhibit modal structures near the root plane. The forcing mode appears near and upstream of the wing, while the response mode develops downstream in the wake. The modal structures for the primary forcing and response modes are aligned with the wingspan, with the response mode similar to the unsteady vortices revealed from DNS. At this frequency, the secondary modes are comprised of spanwise-aligned vortices near the root plane, similar to the primary modes. 
	
	As we increase the frequency, the resolvent gains decay in magnitude and $\sigma_1$ decays faster than $\sigma_2$. Their magnitudes become approximately the same at $St = 0.16$. At this frequency, the spatial characteristics of the primary and secondary forcing-response mode pairs exhibit distinct behavior. The primary forcing and response modes are aligned with  the wingspan and near the root plane, similar to those at lower $St$. The secondary modes, however, exhibit modal structures near the wing tip, in contrast to the secondary modes at lower frequencies which reside near the wing root. 
	
	For $St = 0.18$, the primary forcing-response mode pair appears near the wing tip, while the secondary mode pair develops at the root plane. Such behavior persists as we increase the frequency to $St = 0.20$.  For $St \ge 0.18$, primary modes are tip-dominated while secondary modes are root-dominated around this wing. This means that root and wing tip modes switch their order of amplification at $St \approx 0.18$, i.e., mode switching.
	
	\begin{figure}
		\centering
		\begin{tikzpicture}
		\node[anchor=south west,inner sep=0] (image) at (0,0) {\includegraphics[trim=0mm 0mm 0mm 0mm, clip,width=1.0\textwidth]{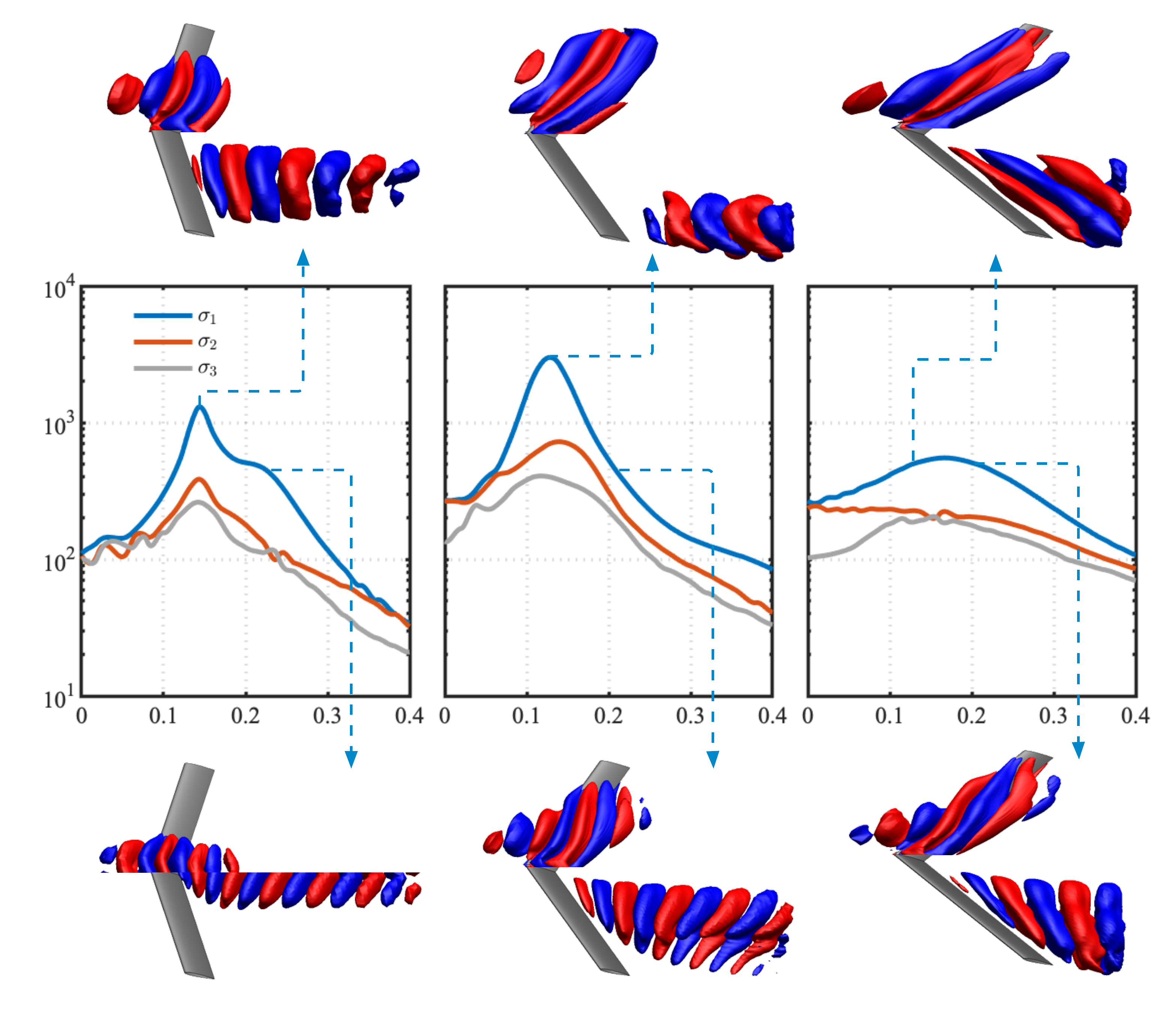}};
		\scriptsize
		\node[align=left] at (1.2,11.2) {\rotatebox{0}{\color{matblue}$\hat{f}_1$}};
		\node[align=left] at (1.2,9.4) {\rotatebox{0}{\color{matblue}$\hat{q}_1$}};
		\node[align=left] at (1.2,2.6) {\rotatebox{0}{\color{matblue}$\hat{f}_1$}};
		\node[align=left] at (1.2,0.8) {\rotatebox{0}{\color{matblue}$\hat{q}_1$}};
		\node[align=left] at (5.6,11.2) {\rotatebox{0}{\color{matblue}$\hat{f}_1$}};
		\node[align=left] at (5.6,9.4) {\rotatebox{0}{\color{matblue}$\hat{q}_1$}};
		\node[align=left] at (5.6,2.6) {\rotatebox{0}{\color{matblue}$\hat{f}_1$}};
		\node[align=left] at (5.6,0.8) {\rotatebox{0}{\color{matblue}$\hat{q}_1$}};
		\node[align=left] at (10.0,11.2) {\rotatebox{0}{\color{matblue}$\hat{f}_1$}};
		\node[align=left] at (10.0,9.4) {\rotatebox{0}{\color{matblue}$\hat{q}_1$}};
		\node[align=left] at (10.0,2.6) {\rotatebox{0}{\color{matblue}$\hat{f}_1$}};
		\node[align=left] at (10.0,0.8) {\rotatebox{0}{\color{matblue}$\hat{q}_1$}};
		\node[align=left] at (2.8,3.2) {\rotatebox{0}{$St$}};
		\node[align=left] at (7.0,3.2) {\rotatebox{0}{$St$}};
		\node[align=left] at (11.20,3.2) {\rotatebox{0}{$St$}};
		\node[align=left] at (0.2,6.2) {\rotatebox{90}{$\sigma$}};
		\node[align=left] at (1.45,8.65) {\rotatebox{0}{$\Lambda = 15^\circ$}};
		\node[align=left] at (5.60,8.65) {\rotatebox{0}{$\Lambda = 30^\circ$}};
		\node[align=left] at (9.75,8.65) {\rotatebox{0}{$\Lambda = 45^\circ$}};
		\end{tikzpicture} \vspace{-8mm}
		\caption{Resolvent gain distribution for the top three mode pairs and forcing-response mode pairs for selected frequencies for $(sAR,\alpha) = (4,20^\circ)$ and $\Lambda =  15^\circ$, $30^\circ$, and $45^\circ$. Primary forcing ($\hat{f}$) and response ($\hat{q}$) modes shown with isosurfaces of velocity $\hat{u}_y \in [-0.2,0.2]$.} 
		\label{fig:gainCurves_sweep}
	\end{figure}
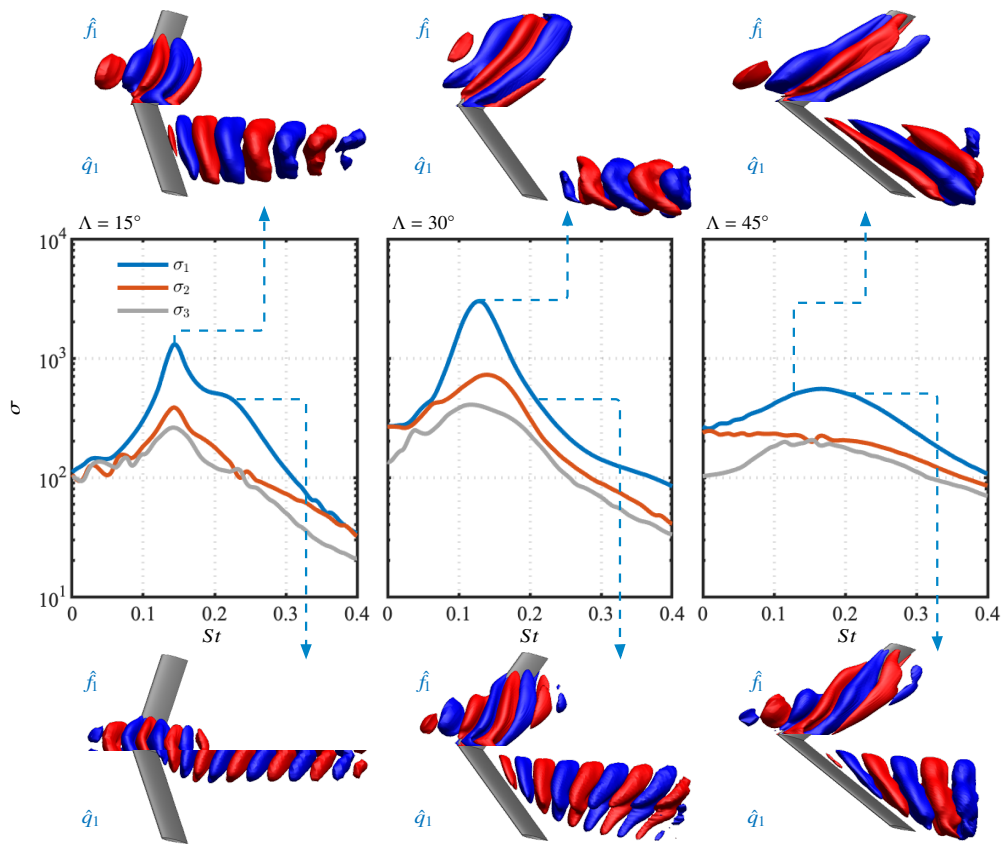
	
	The mode switching phenomenon is also observed for swept wings with $\Lambda = 15^\circ$. For such wings, root-supported structures appear as the primary forcing-response pairs at $St = 0.14$, as shown in figure \ref{fig:gainCurves_sweep} (left). A distinct mode switching is observed over this wing, as the forcing-response pairs gradually transition towards the root at $z/c \approx 0$ with the increase in $St$. This type of concentrated resolvent mode at the wing root also appears for the unswept wings at $St = 0.20$ as a secondary mode, as shown in figure \ref{fig:modeSwitching} (bottom, right).
	
	For $\Lambda = 30^\circ$, mode switching also occurs towards the root with the increase in $St$, in an opposite trend to the unswept wings. The dominant response modes at lower frequencies appear at the wing tip, as shown in figure \ref{fig:gainCurves_sweep} (middle). There is a gradual transition to root-supported modes as the frequency increases. At a higher sweep angle, $\Lambda = 45^\circ$, no mode switching occurs. The region of dominance of the forcing and response modes is slightly invariant for the frequencies shown herein.
			
	In contrast with the lower sweep angle wings, for $\Lambda = 45^\circ$, forcing and response modes are dominant at distinct wingspan locations, as shown in figure \ref{fig:gainCurves_sweep} (right). Response modes are tip-dominated while forcing structures appear upstream near the root plane, extending over the wingspan aligned with the sweep angle. For all $(sAR,\alpha) = (4,20^\circ)$ wings, the highest amplification is found for $\Lambda = 30^\circ$, at $St \approx 0.12$. At $\Lambda = 45^\circ$, the dominant gain is an order of magnitude lower. This finding suggests that it is challenging to perturb flows over $\Lambda = 45^\circ$ wings. These wings are steady because self-sustained flow disturbances cannot introduce sufficient energy into the wake to generate vortex shedding. 
	
	\subsection{Resolvent wavemakers}
	\label{sec:results_wavemakers}
	To characterize the self-sustained unsteadiness in the flows over swept wings, we study the spatial overlap between the forcing and response modes that supports the continuous formation of vortical structures.  Since the forcing modes show regions receptive to external perturbations and the response modes reveal the structures being excited due to the forcing, the region over which forcing and response modes overlap can be interpreted as a mechanism for self-sustained oscillations in the flow.  This idea is similar to the wavemaker concept deduced from direct and adjoint eigenmodes presented in \citet{Giannetti:JFM07}.
	
	Through the wavemaker analysis, previous studies identified critical points responsible for sustaining wake shedding on laminar wakes around cylinders \citep{Strykowski:90,Hill:AIAA92} and regions associated with their primary and secondary instability modes \citep{Giannetti:JFM07,Giannetti:JFM10}. Moreover, wavemakers revealed the physical mechanisms responsible for tonal noise generation in high-Reynolds number flows over airfoils \citep{FosasdePando:JFM17} and self-sustained flow instabilities in transonic buffet regimes \citep{Paladini:PRF19}.
	
	In the aforementioned studies, wavemakers were derived from direct and adjoint global stability eigenmodes.  Our formulation derives wavemakers from global resolvent modes and it is closely related to the structural sensitivity devised by \cite{Qadri:PRF17} and to the resolvent wavemaker studied by \cite{Skene:JFM22}. The present resolvent wavemaker is not identical to the eigenvector-based wavemaker. Using the time-averaged base flow, the present forcing terms encapsulate nonlinear effects as a internal feedback mechanism within the flow. Hence, the spatial overlap between forcing and response identifies regions responsible for self-sustained wake oscillations. 
	
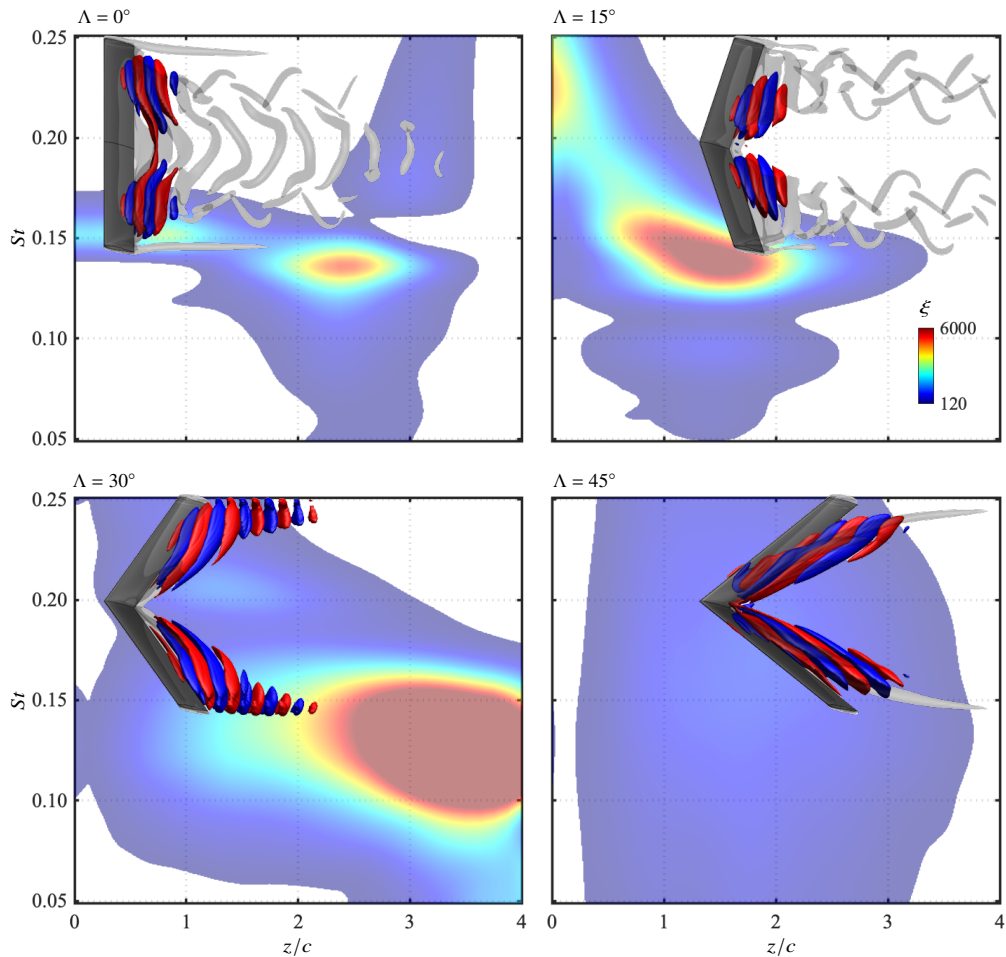
\begin{figure}
		\centering
		\begin{tikzpicture}
		\node[anchor=south west,inner sep=0] (image) at (0,0) {\includegraphics[trim=0mm 0mm 0mm 0mm, clip,width=1.0\textwidth]{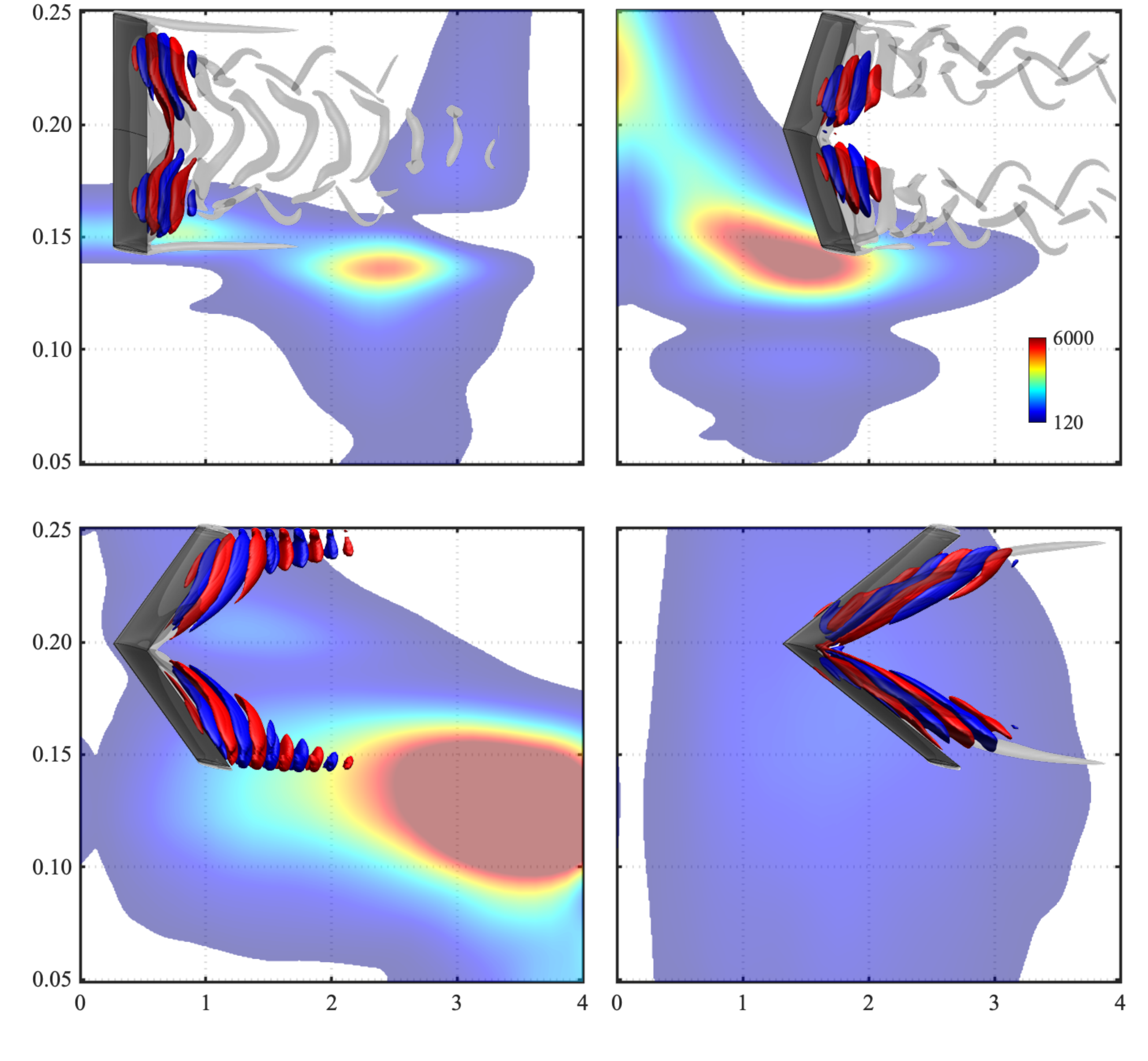}};
		\scriptsize
		\node[align=left] at (12.2,8.75) {\rotatebox{0}{$\xi$}};
		\node[align=left] at (0.2,3.6) {\rotatebox{90}{$St$}};
		\node[align=left] at (0.2,9.7) {\rotatebox{90}{$St$}};
		\node[align=left] at (3.9,0.35) {\rotatebox{0}{$z / c$}};
		\node[align=left] at (10.3,0.35) {\rotatebox{0}{$z / c$}};
		\node[align=left] at (1.35,12.60) {\rotatebox{0}{$\Lambda = 0^\circ$}};
		\node[align=left] at (7.70,12.60) {\rotatebox{0}{$\Lambda = 15^\circ$}};
		\node[align=left] at (1.35,6.50) {\rotatebox{0}{$\Lambda = 30^\circ$}};
		\node[align=left] at (7.70,6.50) {\rotatebox{0}{$\Lambda = 45^\circ$}};
		\end{tikzpicture} \vspace{-8mm}
		\caption{Wingspan location of primary resolvent wavemakers with isocontours of $\xi$ for $0.05 \le St \le 0.25$ for $(sAR,\alpha) = (4,20^\circ)$ wings with $0^\circ \le \Lambda \le 45^\circ$.  Resolvent wavemaker modes at $St = 0.14$ are shown with isosurfaces of $\hat{u}_y / \| \hat{u}_y \|_\infty = \pm 0.1$ and instantaneous $Q=1$ are gray-colored isosurfaces. } 
		\label{fig:wavemakers}
	\end{figure}

	Herein, the resolvent wavemaker modes are directly obtained from the resolvent modes, as the Hadamard product of forcing~and~response modes
	\begin{equation}
	\hat{\mathbi{w}} =  \hat{\mathbi{f}} \circ \hat{\mathbi{q}} \mbox{   ,}
	\label{eq:wi}
	\end{equation}
	where $\hat{\mathbi{w}}$ is the resolvent wavemaker mode. The resolvent modes presented herein are defined with the five  state variables, $\hat{\mathbi{f}} = [\hat{\mathbi{f}}_\rho,\hat{\mathbi{f}}_{u_x},\hat{\mathbi{f}}_{u_y},\hat{\mathbi{f}}_{u_z},\hat{\mathbi{f}}_T]$ and $\hat{\mathbi{q}} = [\hat{\mathbi{q}}_\rho,\hat{\mathbi{q}}_{u_x},\hat{\mathbi{q}}_{u_y},\hat{\mathbi{q}}_{u_z},\hat{\mathbi{q}}_T]$. We define our resolvent wavemaker gain $\xi$ as 
	\begin{equation}
	\xi = \sigma^2 \int_{S} |\hat{\mathbi{w}}(\mathbi{x})| \ {\rm d} S \mbox{   ,}
	\label{eq:xi}
	\end{equation}
	which follows $\xi = \sigma^2 | \langle \hat{\mathbi{f}},\hat{\mathbi{q}} \rangle |$, derived by \cite{Skene:JFM22}.  This is similar to the one presented in \cite{Ribeiro:JFM22}. Qualitatively, both definitions of $\xi$ result in similar discussions and interpretations. The present expression for the resolvent wavemaker gain provides a proper quantitative definition \citep{Skene:JFM22}. The resolvent wavemaker gain $\xi$ can also be computed for each spanwise slice and each frequency. To this end, we consider $S = S(x,y)$, as $z$-normal planes at different spanwise locations, to build the $\xi$-contours shown in figure \ref{fig:wavemakers}. Through this analysis, we highlight the spatial support of the resolvent wavemaker over~the~wingspan.
	
	Let us focus our resolvent wavemaker analysis on the flow over the unswept wing with $(sAR,\alpha,\Lambda) = (4,20^\circ,0^\circ)$, as shown in figure \ref{fig:wavemakers} (top, left). At $St=0.14$,  triglobal resolvent wavemakers with high $\xi$ appear between $2 \lesssim z/c \lesssim 3$ in the near wake. The resolvent wavemakers at this region support the formation of unsteady root vortices that propagate downstream in the wake. This resolvent wavemaker region is also characterized by the formation of braid-like structures that connect to the root shedding as vortex loops \citep{Zhang:JFM20b}. Resolvent wavemakers for $(sAR,\alpha,\Lambda) = (4,20^\circ,15^\circ)$ also show similar shedding behavior, as seen in figure \ref{fig:wavemakers} (top, right).
	
	Resolvent wavemakers are also revealed for steady flows. The overlap of forcing and response modes for flows over wings with $(sAR,\alpha,\Lambda) = (4,20^\circ,30^\circ)$, as shown in figure \ref{fig:wavemakers} (bottom, left), develops over the wing and extends into the wake aligned with the wing tip. These resolvent wavemakers extend over the entire wingspan, being stronger and larger than the ones exhibited around wings with lower sweep angles. The $\xi$ peak appears at the wing tip at $St = 0.12$, indicating that the tip region is more susceptible to develop unsteadiness around this wing.
	
	For $\Lambda = 45^\circ$, shown in figure \ref{fig:wavemakers} (bottom, right), we reveal that resolvent wavemakers emerge from the leading edge near the root plane towards the wing tip and downstream at the wake, overlapping the region where steady ram-horn-shaped vortices appear in the DNS.  These resolvent wavemakers exhibit a region of the flow field with high receptiveness to amplify forcing structures and disturb the steady ram-horn vortex. Because the dominant resolvent wavemaker around the $\Lambda = 45^\circ$ wing have a low $\xi$, in spite of occupying a large region of the wake, the energy they introduce to the flow field is insufficient to disturb the wake.
	
	The resolvent wavemakers further exhibit the root- and tip-dominated modal characteristics and the mode switching phenomenon in figure \ref{fig:wavemakers}, in agreement with the forcing-response modal behavior shown in figures \ref{fig:modeSwitching} and \ref{fig:gainCurves_sweep}. For instance, the resolvent wavemaker modes at the peak $\xi$ values for the unswept wing appear near $z/c \approx 2$, with a gradual transition from root-supported to tip-dominated modes as $St$ increases. Moreover, for the $\Lambda = 15^\circ$ wing, there is a transition in the dominant region of resolvent wavemaker support from $z/c \approx 1.5$ at lower frequencies to $z/c \approx 0$ at higher frequencies, as shown in figure \ref{fig:wavemakers} (top, right). Lastly, for the $\Lambda = 30^\circ$ wing, there is a tip-to-root transition with the increase in $St$ while the peak resolvent wavemakers for $\Lambda = 45^\circ$ are invariant over the frequencies, appearing near $z/c \approx 2$.
	
	\subsection{Forcing-to-response dynamics}
	\label{sec:results_sweep}
	
\begin{figure}
		\centering
		\begin{tikzpicture}
		\node[anchor=south west,inner sep=0] (image) at (0,0) {\includegraphics[trim=5mm 0mm 5mm 0mm, clip,width=1.0\textwidth]{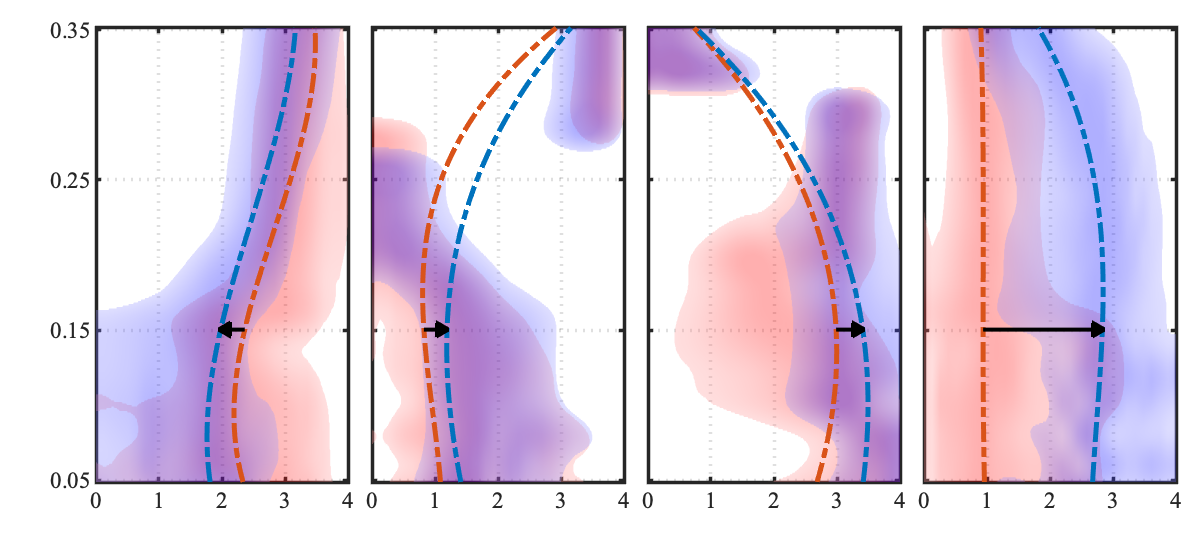}};
		\scriptsize
		\node[align=left] at (0.2,3.6) {\rotatebox{90}{$St$}};
		\node[align=left] at (2.4,0.35) {\rotatebox{0}{$z / c$}};
		\node[align=left] at (5.6,0.35) {\rotatebox{0}{$z / c$}};
		\node[align=left] at (8.8,0.35) {\rotatebox{0}{$z / c$}};
		\node[align=left] at (12.0,0.35) {\rotatebox{0}{$z / c$}};
		\scriptsize
		\node[align=left] at (1.35,6.40) {\rotatebox{0}{$\Lambda = 0^\circ$}};
		\node[align=left] at (4.60,6.40) {\rotatebox{0}{$\Lambda = 15^\circ$}};
		\node[align=left] at (7.80,6.40) {\rotatebox{0}{$\Lambda = 30^\circ$}};
		\node[align=left] at (10.95,6.40) {\rotatebox{0}{$\Lambda = 45^\circ$}};
		\end{tikzpicture} \vspace{-7mm}
		\caption{Wingspan locations of dominant forcing (red) and response (blue) with contours of $\mathbi{\Omega}_{\hat{\mathbi{f}}}$ and $\mathbi{\Omega}_{\hat{\mathbi{q}}} \in  [0.4,1.0]$ for $0.05 \le St \le 0.35$ for $sAR = 4$ wings with $\alpha = 30^\circ$ with $0^\circ \le \Lambda \le 45^\circ$. Dot-dashed lines are polynomial fit of maximum $z/c$ of forcing and response at each $St$. Arrows show direction of optimal forcing-to-response at $St = 0.15$.} 
		\label{fig:oblique_sweep}
	\end{figure}
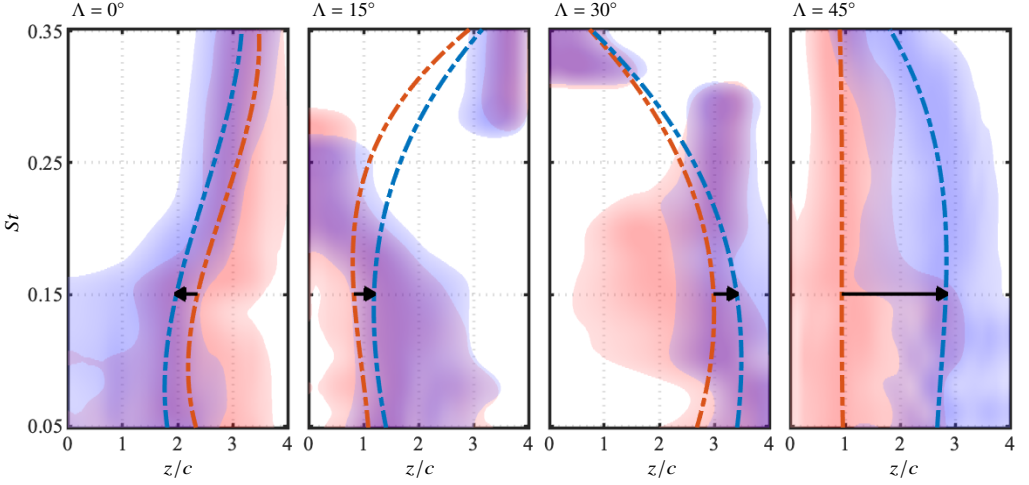

	Let us further explain how perturbations emerge around swept wings, by analyzing the overlap between the forcing and response modes in the spanwise direction. To this end, we  integrate the norm of $\hat{\mathbi{f}}$ and $\hat{\mathbi{q}}$ over $z$-normal planes, analogous to the resolvent wavemaker modes analysis, as
	\begin{equation}
	\mathbi{\Omega}_{\hat{\mathbi{f}}}(z) = \int_{S(x,y)} \|\hat{\mathbi{f}}\|_2 \ {\rm d} S \quad \text{and} \quad \mathbi{\Omega}_{\hat{\mathbi{q}}}(z) = \int_{S(x,y)} \|\hat{\mathbi{q}}\|_2 \ {\rm d} S \mbox{   ,}
	\label{eq:integralF}
	\end{equation}
	where $\|\hat{\mathbi{f}}\|_2$ and $\|\hat{\mathbi{q}}\|_2$ are the $2$-norm of $[\hat{\mathbi{f}}_{\rho},\hat{\mathbi{f}}_{u_x},\hat{\mathbi{f}}_{u_y},\hat{\mathbi{f}}_{u_z},\hat{\mathbi{f}}_{T}]$ and $[\hat{\mathbi{q}}_{\rho},\hat{\mathbi{q}}_{u_x},\hat{\mathbi{q}}_{u_y},\hat{\mathbi{q}}_{u_z},\hat{\mathbi{q}}_{T}]$, respectively, at each grid point of the computational domain. By performing the integral over $S(x,y)$, we obtain $\mathbi{\Omega}_{\hat{\mathbi{f}}}$ and $\mathbi{\Omega}_{\hat{\mathbi{q}}}$ computed for each spanwise slice and for each frequency. Here, we plot their contours normalized by the maximum $\mathbi{\Omega}_{\hat{\mathbi{f}}}$ and $\mathbi{\Omega}_{\hat{\mathbi{q}}}$ at each $St$, to emphasize the spatial support of forcing and response over the wingspan, as shown in figure \ref{fig:oblique_sweep} for wings at $\alpha = 30^\circ$ with $0^\circ \le \Lambda \le 45^\circ$ . The locations of the maximum strength of forcing and response modes are shown by the dot-dashed lines. Black arrows indicate the direction from the maximum forcing to the maximum response at $St = 0.15$. This analysis depicts the preferential direction in which optimal forcing is transferred to optimal response over the wingspan at each frequency.
	
	For unswept wings, shown in figure \ref{fig:oblique_sweep} (left), the optimal forcing structures appear closer to the wing tip than the response modes, which are slightly shifted towards the root, suggesting that fluctuations are directed towards the root. Indeed, as seen in the DNS, unsteadiness is concentrated towards the root, as evident from figure \ref{fig:validation}, also in agreement with the results reported by \cite{Zhang:JFM20}. In addition, the flow around the wing tip for unswept wings is characterized by an almost steady tip vortex, suggesting that it is likely hard to amplify flow oscillations near the tip.
	
	For swept wings, fluctuations are directed towards the wing tip. For $\Lambda = 15^\circ$, both forcing and response modes appear near the wing root. At the vortex shedding frequency for this wing, $St \approx 0.15$, we observe forcing and response  modes to be dominant at $z/c \approx 1$, with the forcing mode supported closer to the wing root than the response mode.  This concurs with the flow field we observe in the DNS, as vortices are formed near the wing root and evolve towards the wing tip where spanwise vortices appear and propagate in the wake. For the  $\Lambda = 30^\circ$ wing, the dominant forcing-response mode pair emerges near the wing tip at low $St$, as seen in figure \ref{fig:oblique_sweep}. For this wing, low-frequency vortical structures emerge downstream in the wake aligned at the tip, as shown in figure \ref{fig:validation}. 
	
	For the $\Lambda = 45^\circ$ wing, the distance between the maximum forcing and response mode locations significantly increases. For this sweep angle, the region of forcing is centered at $z/c \approx 1$, while the response is supported mostly at $z/c \approx 3$. As the peak $\sigma_1$ is lower for this wing compared to lower sweep angle planforms, we argue that a significant amount of energy is required for an external forcing to perturb the wakes of highly swept wings. For all wings with $sAR=4$, this distance between the dominant forcing-response mode pairs is strongly associated with the sweep angle, while having a minor dependency on the angle of attack and presenting a gradual decrease with the frequency.
	
	The direction from forcing-to-response revealed by the optimal triglobal resolvent modes suggests a spanwise advection of flow structures associated with the sweep angle. As shown previously, we can relate the forcing-to-response characteristics to the vortical fluctuations observed in the DNS. We can further relate these findings to the modal convective speed from  biglobal stability analysis over swept wings \citep{Crouch:JFM19,Paladini:PRF19,Plante:JFM21}. Triglobal resolvent modes also reveal the advection of perturbations over the wingspan related to the sweep angle, the attenuation of flow unsteadiness, and the resilience to amplify perturbations at high sweep angles. Even for unswept wings, the triglobal analysis uncovers a preferential root-direction for advection~of~oscillations.
	
	\subsection{Influence of the aspect ratio}
	\label{sec:results_ar}
	
	High sweep angle and low aspect ratio restrict the emergence of fluctuations in flows over finite wings. As shown in figure \ref{fig:oblique_sweep}, tip- and root-dominated modes may extend over $1$ or $2$ chord-lengths over the wingspan. For this reason, for flows over wings with $sAR < 2$, the dominance of the global modes may not be associated with root or tip regions, as they extend over the entire wingspan.  
	
	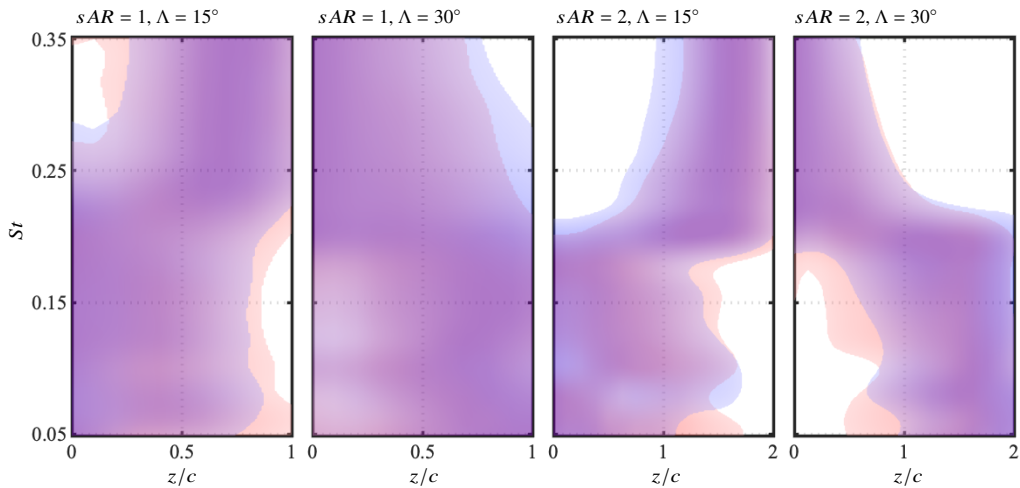
\begin{figure}
		\centering
		\begin{tikzpicture}
		\node[anchor=south west,inner sep=0] (image) at (0,0) {\includegraphics[trim=5mm 0mm 5mm 0mm, clip,width=1.0\textwidth]{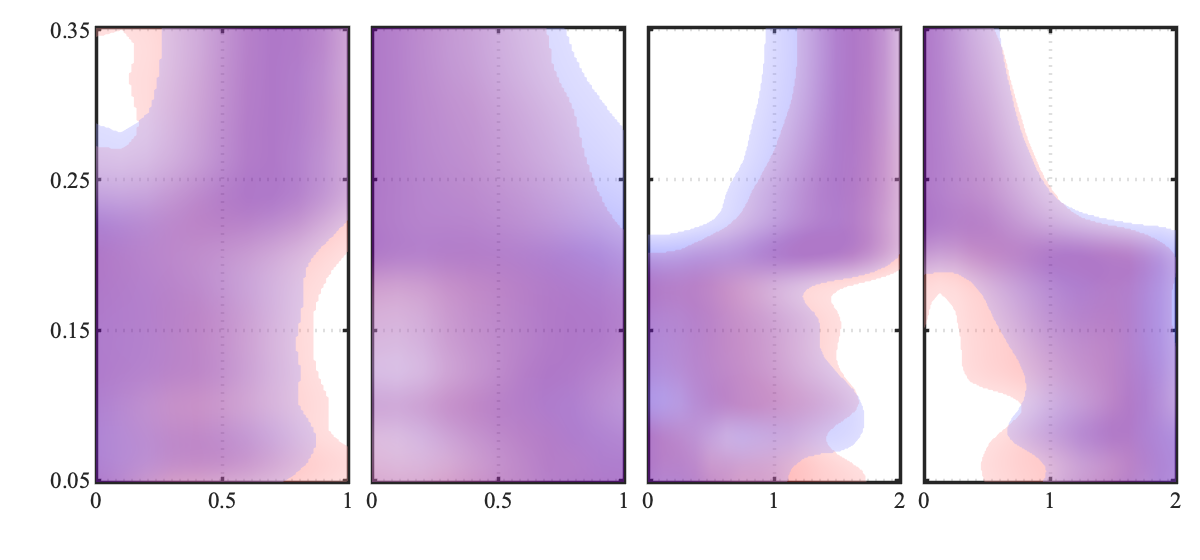}};
		\scriptsize
		\node[align=left] at (0.2,3.6) {\rotatebox{90}{$St$}};
		\node[align=left] at (2.4,0.35) {\rotatebox{0}{$z / c$}};
		\node[align=left] at (5.6,0.35) {\rotatebox{0}{$z / c$}};
		\node[align=left] at (8.8,0.35) {\rotatebox{0}{$z / c$}};
		\node[align=left] at (12.0,0.35) {\rotatebox{0}{$z / c$}};
		\scriptsize
		\node[align=left] at (1.95,6.40) {\rotatebox{0}{$sAR = 1$, $\Lambda = 15^\circ$}};
		\node[align=left] at (5.15,6.40) {\rotatebox{0}{$sAR = 1$, $\Lambda = 30^\circ$}};
		\node[align=left] at (8.30,6.40) {\rotatebox{0}{$sAR = 2$, $\Lambda = 15^\circ$}};
		\node[align=left] at (11.45,6.40) {\rotatebox{0}{$sAR = 2$, $\Lambda = 30^\circ$}};
		\end{tikzpicture} \vspace{-7mm}
		\caption{Wingspan locations of dominant forcing (red) and response (blue) with contours of $\mathbi{\Omega}_{\hat{\mathbi{f}}}$ and $\mathbi{\Omega}_{\hat{\mathbi{q}}} \in  [0.4,1.0]$ for $0.05 \le St \le 0.35$ for wings at $\alpha = 30^\circ$, $sAR = 2$ and $1$, and $\Lambda = 15^\circ$ and $30^\circ$.} 
		\label{fig:arEffect}
	\end{figure}

	For flows over  $sAR = 2$ wings, we observe a gradual transition between root-dominated and tip-dominated forcing and response modes, as shown in figure \ref{fig:arEffect}. For $\Lambda = 15^\circ$, the optimal forcing-response mode pair appears near the root for lower frequencies and at the wing tip for higher frequencies,  characterizing a root-to-tip mode switching. For $\Lambda = 30^\circ$, the trend is opposite, with wing tip modes at lower frequencies and root modes at higher frequencies, characterizing a tip-to-root mode switching. These features are similar to the mode switching observed for these sweep angles with $sAR = 4$, as shown in figure \ref{fig:oblique_sweep}.  

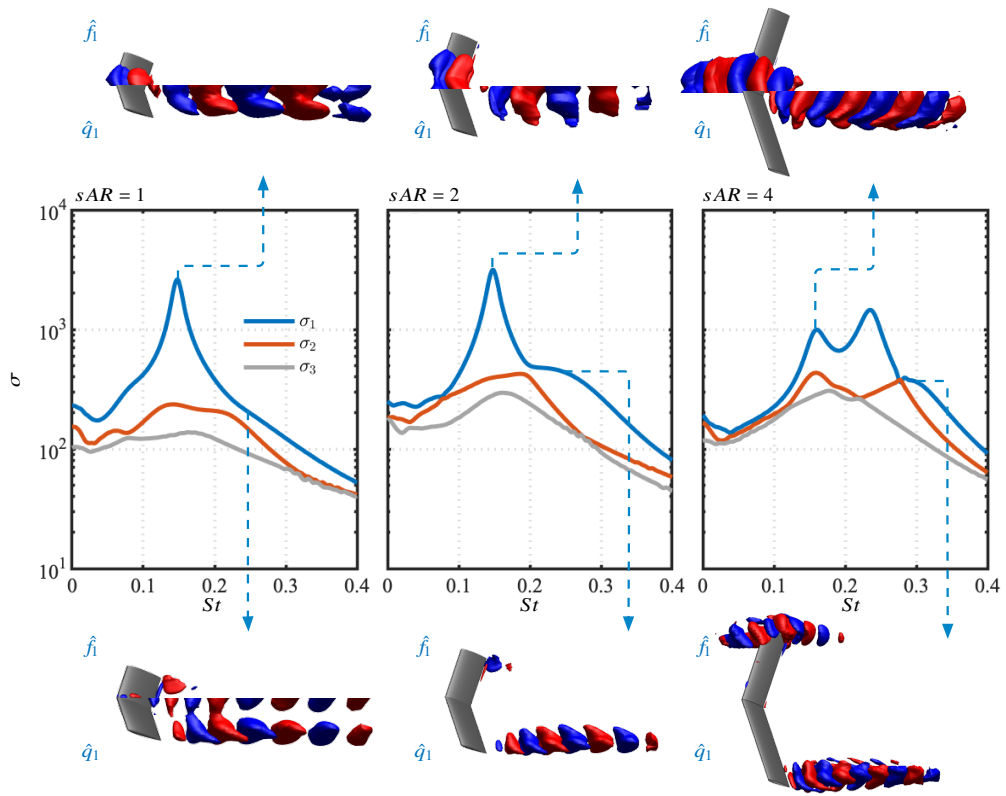
\begin{figure}
		\centering
		\begin{tikzpicture}
		\node[anchor=south west,inner sep=0] (image) at (0,0) {\includegraphics[trim=0mm 0mm 0mm 0mm, clip,width=1.0\textwidth]{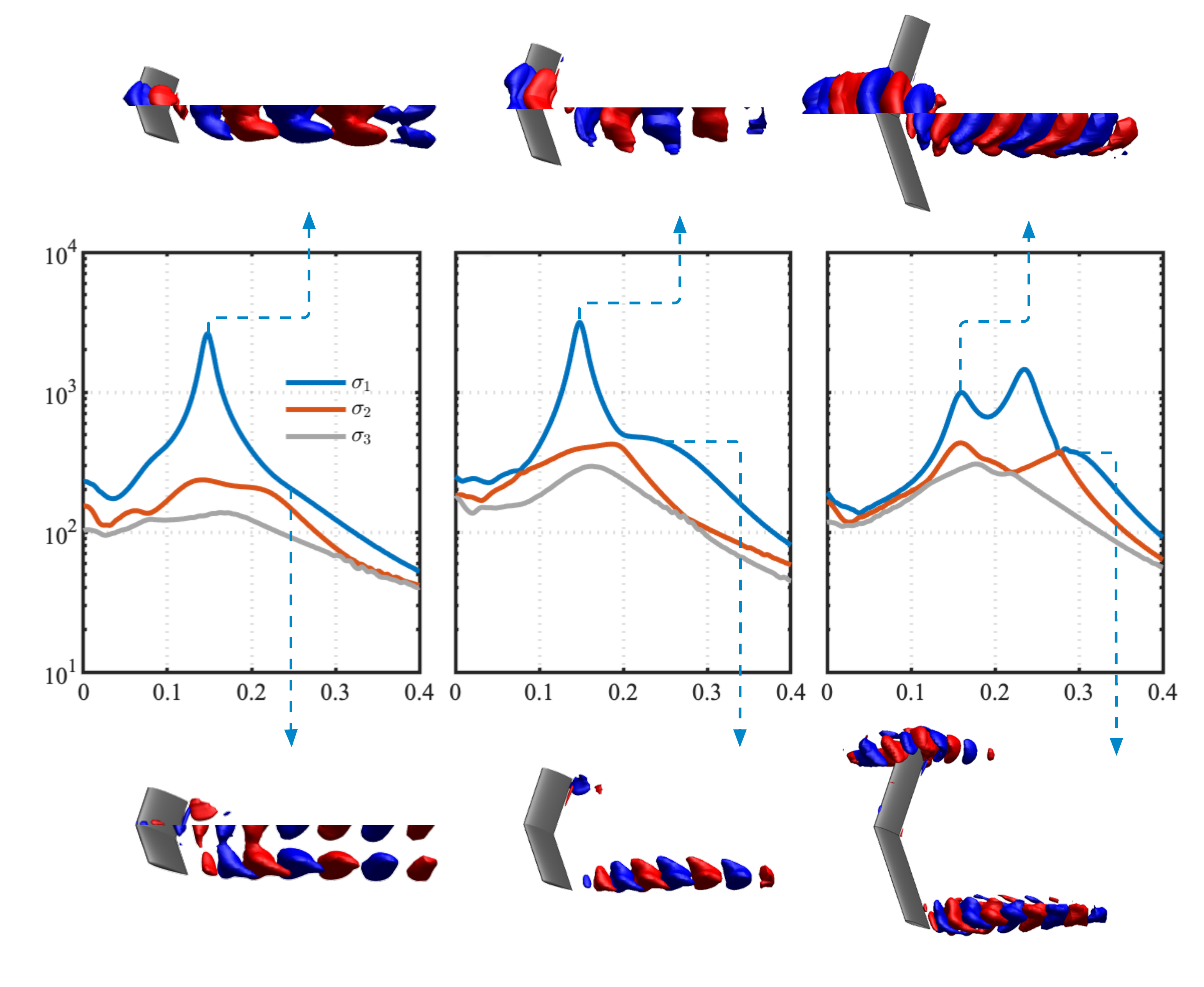}};
		\scriptsize
		\node[align=left] at (1.2,10.6) {\rotatebox{0}{\color{matblue}$\hat{f}_1$}};
		\node[align=left] at (1.2,9.3) {\rotatebox{0}{\color{matblue}$\hat{q}_1$}};
		\node[align=left] at (1.2,2.5) {\rotatebox{0}{\color{matblue}$\hat{f}_1$}};
		\node[align=left] at (1.2,1.1) {\rotatebox{0}{\color{matblue}$\hat{q}_1$}};
		\node[align=left] at (5.6,10.6) {\rotatebox{0}{\color{matblue}$\hat{f}_1$}};
		\node[align=left] at (5.6,9.3) {\rotatebox{0}{\color{matblue}$\hat{q}_1$}};
		\node[align=left] at (5.6,2.5) {\rotatebox{0}{\color{matblue}$\hat{f}_1$}};
		\node[align=left] at (5.6,1.1) {\rotatebox{0}{\color{matblue}$\hat{q}_1$}};
		\node[align=left] at (9.3,10.6) {\rotatebox{0}{\color{matblue}$\hat{f}_1$}};
		\node[align=left] at (9.3,9.3) {\rotatebox{0}{\color{matblue}$\hat{q}_1$}};
		\node[align=left] at (9.3,2.5) {\rotatebox{0}{\color{matblue}$\hat{f}_1$}};
		\node[align=left] at (9.3,1.1) {\rotatebox{0}{\color{matblue}$\hat{q}_1$}};
		\node[align=left] at (1.45,8.45) {\rotatebox{0}{$sAR = 1$}};
		\node[align=left] at (5.60,8.45) {\rotatebox{0}{$sAR = 2$}};
		\node[align=left] at (9.75,8.45) {\rotatebox{0}{$sAR = 4$}};
		\node[align=left] at (0.2,6.0) {\rotatebox{90}{$\sigma$}};
		\node[align=left] at (2.8,3.05) {\rotatebox{0}{$St$}};
		\node[align=left] at (7.0,3.05) {\rotatebox{0}{$St$}};
		\node[align=left] at (11.25,3.05) {\rotatebox{0}{$St$}};
		\end{tikzpicture} \vspace{-8mm}
		\caption{Resolvent gain distribution and forcing-response mode pairs over frequency for $(\alpha,\Lambda) = (30^\circ,15^\circ)$ and $1 \le sAR \le 4$. Forcing ($\hat{f}$) and response ($\hat{q}$) modes shown with isosurfaces of velocity $\hat{u}_y \in [-0.2,0.2]$. Mode switching is absent for $sAR = 1$ due to merging of root and wingtip perturbations on the wake.} 
		\label{fig:AR_sw15}
	\end{figure}

	For wings with a low aspect ratio, the growth of root-dominated and tip-dominated perturbations is constrained and mode switching does not occur for $sAR = 1$, as shown in figure \ref{fig:AR_sw15}. Distinguishing between root-dominated and tip-dominated modes may be challenging for flows over $sAR = 1$ wings as forcing and response mode pairs appear globally, extending over the entire wingspan, independent of the sweep angle. Therefore, flows around wings with $sAR = 1$ tend to exhibit similar wake characteristics over different sweep angles. Indeed, the wake patterns for flows over $sAR = 1$ wings at a particular angle of attack and sweep exhibit different characteristics from the flows over higher aspect ratio wings, e.g., $sAR = 2$ and $4$. 
	
	For high-aspect-ratio wings, for instance, the flow around $(sAR,\alpha,\Lambda) = (4,30^\circ,15^\circ)$ wings, we observe in the DNS that the wake shedding structures appear over the entire wingspan. The resolvent modes depict these structures in three different flow mechanisms. As shown in figure \ref{fig:AR_sw15} (right) for $sAR = 4$, there are two types of root-dominated modes, which were also previously identified for this wing at $\alpha = 20^\circ$, shown in figure \ref{fig:gainCurves_sweep} (left). The first one is characterized by root-dominated structures and appears at $St = 0.15$, while the second type, with a high $\sigma_1$, develops at $St = 0.25$ with compact root-concentrated modes. The third type is comprised of tip-dominated modes that become primary as the frequency increases to $St = 0.28$. These modes were primary at $\alpha = 20^\circ$ and $St \approx 0.40$, as shown in figure \ref{fig:gainCurves_sweep}, although for $\alpha = 30^\circ$ they present a higher amplification gain.
	
	For $sAR = 2$, root-dominated modes are primary for $St < 0.20$.  Root-concentrated modes are absent and tip-dominated modes are the primary perturbations for $St \ge 0.20$, as shown in figure \ref{fig:AR_sw15} (middle), characterizing a root-to-tip mode switching. The overall mode switching for $sAR = 2$ is the same, with root-to-tip transition. For $sAR = 1$, shown in figure \ref{fig:AR_sw15} (left), mode switching is absent. Both primary and secondary modes develop over the entire wingspan for all frequencies, as shown for the primary modes at $St = 0.14$. Although mode switching is absent we can still reveal two distinct root- and tip-dominated mechanisms on a single mode over low-aspect-ratio wings. For instance, at $St = 0.24$, modes emerge from the leading edge at the root and from the trailing edge near the tip. Combined, these two types of flow unsteadiness yield a global mode that appears over the entire wingspan.

	\section{Conclusions}
	\label{sec:conclusions}
	We presented the triglobal resolvent analysis of laminar separated flows over swept wings and characterized the effects of wing tip and sweep angle on the wake dynamics. We revealed the forcing and response structures that can be amplified from harmonic oscillations or external actuation over finite wings. In the present triglobal analysis, we have identified the wingspan locations where forcing structures can be amplified near the wing and the regions where the unsteady response develops. We have further characterized the region of dominance of modal structures over the wingspan, with forcing-response mode pairs appearing near the wing root or tip as a function of their characteristic frequency.
	
	Through resolvent wavemakers, we studied the steady to unsteady flow characteristics over swept wings. We showed the regions where self-sustained unsteadiness appear over swept wings and related those to the vortex shedding structures observed in the DNS. We also revealed the most sensitive regions for perturbation amplification in steady wakes over highly swept wings. The forcing-response mode pairs further revealed the mechanisms of spanwise advection of flow structures, which is further related to the spanwise convective speed found for two- and three-dimensional resolvent analysis over swept wings and also associated with the nonlinear flow characteristics observed in DNS.
	
	At last, we showed for low-aspect-ratio wings that localized perturbations with root- or tip-dominant characteristics are limited as modes evolve globally over the entire wingspan. In fact, we have shown that root-and tip-dominant structures can appear over the wing in a single mode for low-aspect-ratio wings. This behavior explains the characteristics of the laminar flows around these wings, as observed in DNS, to be different from the flows over $sAR = 2$ and $4$ wings. These findings provide fundamental insights into future studies on flow separation over swept wings at higher Reynolds numbers, in which a wider spectrum of fluctuations is present.
	
	\appendix
	\section{Choice of resolvent discounting parameter}
	\label{sec:appendixDiscounting}
	
	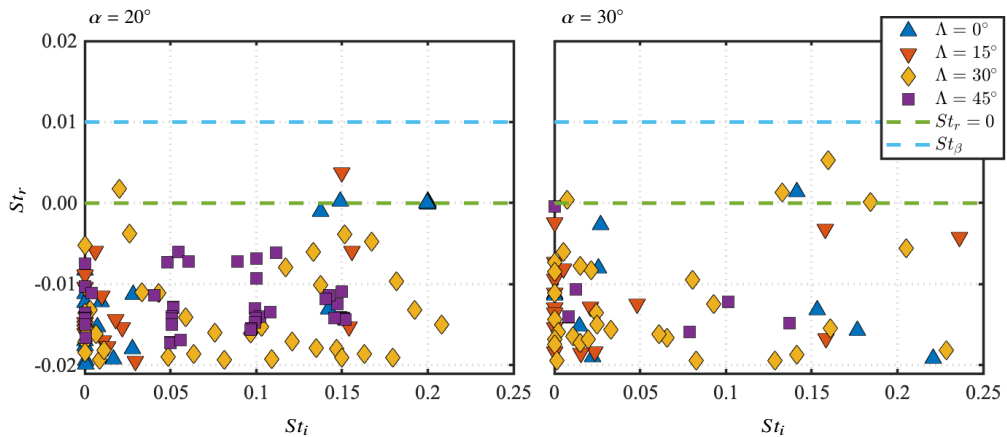
\begin{figure}
		\centering
		\begin{tikzpicture}
		\node[anchor=south west,inner sep=0] (image) at (0,0) {\includegraphics[trim=0mm 0mm 0mm 0mm, clip,width=1.0\textwidth]{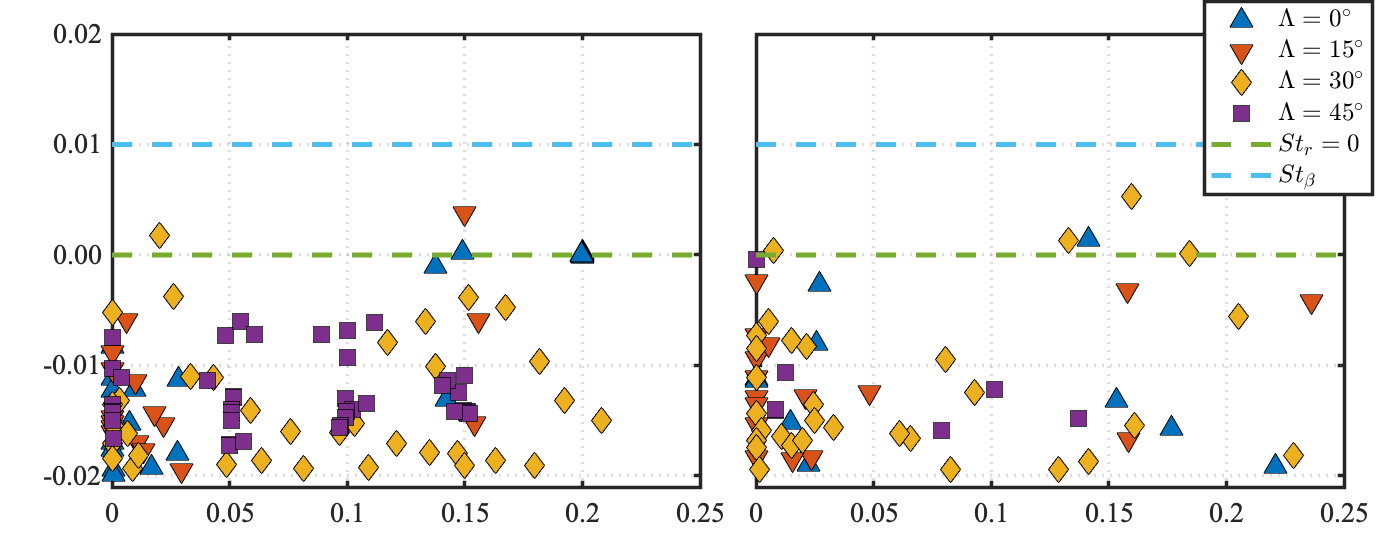}};
		\scriptsize
		\node[align=left] at (0.2,3.00) {\rotatebox{90}{$St_r$}};
		\node[align=left] at (3.90,0.0) {\rotatebox{0}{$St_i$}};
		\node[align=left] at (10.20,0.0) {\rotatebox{0}{$St_i$}};
		\node[align=left] at (1.55,5.40) {\rotatebox{0}{$\alpha= 20^\circ$}};
		\node[align=left] at (7.80,5.40) {\rotatebox{0}{$\alpha = 30^\circ$}};
		\end{tikzpicture} \vspace{-6mm}
		\caption{Eigenvalues of $\mathsfbi{L}_{\bar{\mathbi{q}}}$ for $sAR= 4$ wings with $\alpha = 20^\circ$ and $30^\circ$, and $0 \le \Lambda \le 45^\circ$. Here $St_r$ and $St_i$ represent the growth rate and temporal frequency, respectively. Green-dashed line shows $St_r = 0$. Cyan-dashed line shows where discounting parameter is set.} 
		\label{fig:eigsDiscounting}
	\end{figure}
	
	Prior to performing the resolvent analysis, we examine the stability characteristics of $\mathsfbi{L}_{\bar{\mathbi{q}}}$ and the need of a discounting parameter. For this task we analyze the eigenspectrum of the linear operator with respect to the time-average base flow. Since this base flow is not the equilibrium point, the eigenvalues of $\mathsfbi{L}_{\bar{\mathbi{q}}}$ do not necessarily present stability characteristics. However, eigenvalue properties are needed to enable the examination of the amplification mechanisms of perturbation over the appropriate time scale.  The eigenvalues of $\mathsfbi{L}_{\bar{\mathbi{q}}}$ are computed using the Krylov--Schur method \citep{Stewart:SIAM02} with 128 vectors for the Krylov subspace and a tolerance residual of $10^{-10}$. This analysis reveals eigenvalues $-i\omega = -i\omega_r + \omega_i$, where $\omega_i$ is the growth rate and $\omega_r$ is the temporal frequency. The Strouhal number scaling (equation \ref{eq:strouhal}) is used throughout this study to report $\omega_i$ and $\omega_r$ as $St_i$ and $St_r$, respectively, as shown in figure \ref{fig:eigsDiscounting}. The discounting parameter is defined in a similar manner as $St_\beta = (\beta / 2 \pi) (c\sin \alpha /U_\infty \cos \Lambda)$. In this manner, $St_\beta$ is directly associated with a physical time window $t_\beta = (2 \pi / \beta)$, which is chosen to be shorter than the the time scale associated with the largest $St_r$. As shown in figure \ref{fig:eigsDiscounting} for flows over $sAR= 4$ wings with $\alpha = 20^\circ$ and $30^\circ$, and $0 \le \Lambda \le 45^\circ$, modes which appear with positive $St_r$ (above green-dashed line) are so-called unstable. 
	
	Through the global stability analysis of all $\mathsfbi{L}_{\bar{\mathbi{q}}}$ operators examined in the present study, we observe that eigenvalues of the unstable modes with higher growth rate lie near the stability margin. The discounting parameter $St_\beta$ must be chosen such that $St_\beta > \max(St_r)$ for all linear operators considered.  For $St_\beta = 0.01$, the resolvent discounting is able to encompass all the unstable modes in the present study, as shown in figure \ref{fig:eigsDiscounting} with a cyan-dashed line. This discounting corresponds to a physical time window of $t_\beta (U_\infty \cos \Lambda / c\sin \alpha) = 100$.
	
	\section{Convergence test for resolvent analysis}
	\label{sec:appendixConvergence}
	We document the randomized resolvent analysis computations in table \ref{tab:resolventValidation}, for a selected case of $(sAR,\alpha,\Lambda) = (2,30^\circ,15^\circ)$.  The convergence of randomized SVD algorithm depends on the number of test vectors $k$ used for sketching \citep{Halko:SIAM11,Ribeiro:PRF20}. As our discussions focus on the modal characteristics of the dominant and sub-dominant resolvent modes, we have analyzed their convergence using $k = 5$, $10$, and $20$. The use of $k = 10$ test vectors was shown to be sufficient to guarantee converged leading modes with error smaller than $1\%$ for the leading $5$ resolvent gains.
	
	\renewcommand{\arraystretch}{1.2}
	\begin{table}
		\vspace*{-3mm}
		\centering
		\begin{tabular}{p{0.30in}p{0.01in}p{1.15in}p{0.01in}p{1.15in}p{0.01in}p{1.15in}}
			&& 
			\multicolumn{1}{l}{\hspace{0mm}$k = 5$} && 
		    \multicolumn{1}{l}{\hspace{0mm}$k = 10$} && 
			\multicolumn{1}{l}{\hspace{0mm}$k = 20$} \\
			$\sigma_1$ && $3058.119140625$ && $3058.1840820312$ && $3058.0112304688$\\
			$\sigma_2$ && $395.9677734375$ && $397.50375366211$ && $397.54058837891$\\
			$\sigma_3$ && $283.04278564453$ && $285.14834594727$ && $285.30682373047$\\
			$\sigma_4$ && $172.99363708496$ && $178.36595153809$ && $179.13203430176$\\
			$\sigma_5$ && $131.17066955566$ && $155.90252685547$ && $156.57176208496$\\
		\end{tabular} \vspace*{-2mm}
		\caption{Convergence of the randomized resolvent analysis using $k = 5$, $10$, and $20$ test vectors shown for the five leading singular values with $(sAR,\alpha,\Lambda) = (2,30^\circ,15^\circ)$ at $St = 0.14$. }
		\label{tab:resolventValidation}
	\end{table}
	
	\section*{Acknowledgments}
	\label{sec:acknowledgments}
	We acknowledge support from the US Air Force Office of Scientific Research (program manager: Dr. G. Abate, grant: FA9550-21-1-0174) and the US Army Research Office (program managers: Drs. M. J. Munson and R. Anthenien, grant: W911NF-21-1-0060). We thank C. S. Skene, T.  R. Ricciardi, M. Amitay, V. Theofilis, and P. J. Schmid for enlightening discussions. Computational resources were provided by the High Performance Computing Modernization Program at the US Department of Defense and the Texas Advanced Computing Center. 
	
	\section*{Declaration of interest}
	\label{sec:doi}
	The authors report no conflict of interest.
	
	\bibliography{taira_refs}

\begin{thebibliography}{68}
\expandafter\ifx\csname natexlab\endcsname\relax\def\natexlab#1{#1}\fi
\def\au#1{#1} \def\ed#1{#1} \def\yr#1{#1}\def\at#1{#1}\def\jt#1{\textit{#1}}
  \def\bt#1{#1}\def\bvol#1{\textbf{#1}} \def\vol#1{#1} \def\pg#1{#1}
  \def\publ#1{#1}\def\arxiv#1{#1}\def\org#1{#1}\def\st#1{\textit{#1}}

\bibitem[Amestoy {\em et~al.\/}(2001)Amestoy, Duff, L'Excellent \&
  Koster]{Amestoy:SIAM01}
{\sc \au{Amestoy, P.R.}, \au{Duff, I.~S.}, \au{L'Excellent, J.-Y.} \&
  \au{Koster, J.}} \yr{2001}  \at{A fully asynchronous multifrontal solver
  using distributed dynamic scheduling}.  \jt{SIAM J. Mat. Anal. App.}
  \bvol{23}~(1),  \pg{15--41}.

\bibitem[Anderson(2010)]{Anderson:10}
{\sc \au{Anderson, J.~D.}} \yr{2010} {\em Fundamentals of aerodynamics\/}.
  \publ{McGraw-Hill}.

\bibitem[Barkley {\em et~al.\/}(2008)Barkley, Blackburn \&
  Sherwin]{Barkley:IJNMF08}
{\sc \au{Barkley, D.}, \au{Blackburn, H.~M.} \& \au{Sherwin, S.~J.}} \yr{2008}
  \at{Direct optimal growth analysis for timesteppers}.  \jt{Int. J. Numer.
  Meth. Fluids}  \bvol{57}~(9),  \pg{1435--1458}.

\bibitem[Barthel {\em et~al.\/}(2022)Barthel, Gomez \& McKeon]{Barthel:PRF22}
{\sc \au{Barthel, B.}, \au{Gomez, S.} \& \au{McKeon, B.~J.}} \yr{2022}
  \at{Variational formulation of resolvent analysis}.  \jt{Phys. Rev. Fluids}
  \bvol{7}~(1),  \pg{013905}.

\bibitem[Br{\`e}s {\em et~al.\/}(2017)Br{\`e}s, Ham, Nichols \&
  Lele]{Bres:AIAAJ17}
{\sc \au{Br{\`e}s, G.~A.}, \au{Ham, F.~E.}, \au{Nichols, J.~W.} \& \au{Lele,
  S.~K.}} \yr{2017}  \at{Unstructured large-eddy simulations of supersonic
  jets}.  \jt{AIAA J.}  \bvol{55}~(4),  \pg{1164--1184}.

\bibitem[Buchholz \& Smits(2006)]{Buchholz:JFM06}
{\sc \au{Buchholz, J. H.~J.} \& \au{Smits, A.~J.}} \yr{2006}  \at{On the
  evolution of the wake structure produced by a low-aspect-ratio pitching
  panel}.  \jt{J. Fluid Mech.}  \bvol{546},  \pg{433--443}.

\bibitem[Burtsev {\em et~al.\/}(2022)Burtsev, He, Hayostek, Zhang, Theofilis,
  Taira \& Amitay]{Burtsev:JFM22}
{\sc \au{Burtsev, A.}, \au{He, W.}, \au{Hayostek, S.}, \au{Zhang, K.},
  \au{Theofilis, V.}, \au{Taira, K.} \& \au{Amitay, M.}} \yr{2022}  \at{Linear
  modal instabilities around post-stall swept finite wings at low {R}eynolds
  numbers}.  \jt{J. Fluid Mech.}  \bvol{944},  \pg{A6}.

\bibitem[Chu(1965)]{Chu:Acta65}
{\sc \au{Chu, B.-T.}} \yr{1965}  \at{On the energy transfer to small
  disturbances in fluid flow (part {I})}.  \jt{Acta Mechanica}  \bvol{1}~(3),
  \pg{215--234}.

\bibitem[Crouch {\em et~al.\/}(2019)Crouch, Garbaruk \& Strelets]{Crouch:JFM19}
{\sc \au{Crouch, J.~D.}, \au{Garbaruk, A.} \& \au{Strelets, M.}} \yr{2019}
  \at{Global instability in the onset of transonic-wing buffet}.  \jt{J. Fluid
  Mech.}  \bvol{881},  \pg{3--22}.

\bibitem[Devenport {\em et~al.\/}(1996)Devenport, Rife, Liapis \&
  Follin]{Devenport:JFM96}
{\sc \au{Devenport, W.~J.}, \au{Rife, M.~C.}, \au{Liapis, S.~I.} \& \au{Follin,
  G.~J.}} \yr{1996}  \at{The structure and development of a wing-tip vortex}.
  \jt{J. Fluid Mech.}  \bvol{312},  \pg{67--106}.

\bibitem[Edstrand {\em et~al.\/}(2018{\natexlab{{\em a\/}}})Edstrand, Schmid,
  Taira \& {Cattafesta III}]{Edstrand:JFM18a}
{\sc \au{Edstrand, A.~M.}, \au{Schmid, P.~J.}, \au{Taira, K.} \&
  \au{{Cattafesta III}, L.~N.}} \yr{2018{\natexlab{{\em a\/}}}}  \at{A parallel
  stability analysis of a trailing vortex wake}.  \jt{J. Fluid Mech.}
  \bvol{837},  \pg{858}.

\bibitem[Edstrand {\em et~al.\/}(2018{\natexlab{{\em b\/}}})Edstrand, Sun,
  Schmid, Taira \& Cattafesta]{Edstrand:JFM18b}
{\sc \au{Edstrand, A.~M.}, \au{Sun, Y.}, \au{Schmid, P.~J.}, \au{Taira, K.} \&
  \au{Cattafesta, L.~N.}} \yr{2018{\natexlab{{\em b\/}}}}  \at{Active
  attenuation of a trailing vortex inspired by a parabolized stability
  analysis}.  \jt{J. Fluid Mech.}  \bvol{855},  \pg{R2}.

\bibitem[Fage \& Johansen(1927)]{FageJohanssen:PRSA27}
{\sc \au{Fage, A.} \& \au{Johansen, F.~C.}} \yr{1927}  \at{On the flow of air
  behind an inclined flat plate of infinite span}.  \jt{Proc. R. Soc. Lond. A}
  \bvol{116}~(773),  \pg{170--197}.

\bibitem[Farghadan {\em et~al.\/}(2021)Farghadan, Towne, Martini \&
  Cavalieri]{Farghadan:AIAA21}
{\sc \au{Farghadan, A.}, \au{Towne, A.}, \au{Martini, E.} \& \au{Cavalieri,
  A.}} \yr{2021}  \bt{A randomized time-domain algorithm for efficiently
  computing resolvent modes}.  \publ{AIAA Paper 2021-2896}.

\bibitem[{Fosas de Pando} \& Schmid(2017)]{Fosas:JoT17}
{\sc \au{{Fosas de Pando}, M.} \& \au{Schmid, P.~J.}} \yr{2017}  \at{Optimal
  frequency-response sensitivity of compressible flow over roughness elements}.
   \jt{J. Turb.}  \bvol{18}~(4),  \pg{338--351}.

\bibitem[{Fosas de Pando} {\em et~al.\/}(2017){Fosas de Pando}, Schmid \&
  Sipp]{FosasdePando:JFM17}
{\sc \au{{Fosas de Pando}, M.}, \au{Schmid, P.~J.} \& \au{Sipp, D.}} \yr{2017}
  \at{On the receptivity of aerofoil tonal noise: an adjoint analysis}.  \jt{J.
  Fluid Mech.}  \bvol{812},  \pg{771--791}.

\bibitem[Freund(1997)]{Freund:AIAAJ97}
{\sc \au{Freund, J.~B.}} \yr{1997}  \at{Proposed inflow/outflow boundary
  condition for direct computation of aerodynamic sound}.  \jt{AIAA J.}
  \bvol{35}~(4),  \pg{740--742}.

\bibitem[Giannetti {\em et~al.\/}(2010)Giannetti, Camarri \&
  Luchini]{Giannetti:JFM10}
{\sc \au{Giannetti, F.}, \au{Camarri, S.} \& \au{Luchini, P.}} \yr{2010}
  \at{Structural sensitivity of the secondary instability in the wake of a
  circular cylinder}.  \jt{J. Fluid Mech.}  \bvol{651},  \pg{319--337}.

\bibitem[Giannetti \& Luchini(2007)]{Giannetti:JFM07}
{\sc \au{Giannetti, F.} \& \au{Luchini, P.}} \yr{2007}  \at{Structural
  sensitivity of the first instability of the cylinder wake}.  \jt{J. Fluid
  Mech.}  \bvol{581}~(1),  \pg{167--197}.

\bibitem[G{\'o}mez {\em et~al.\/}(2016)G{\'o}mez, Blackburn, Rudman, Sharma \&
  Mc{K}eon]{Gomez:JFM16}
{\sc \au{G{\'o}mez, F.}, \au{Blackburn, H.~M.}, \au{Rudman, M.}, \au{Sharma,
  A.~S.} \& \au{Mc{K}eon, B.~J.}} \yr{2016}  \at{A reduced-order model of
  three-dimensional unsteady flow in a cavity based on the resolvent operator}.
   \jt{J. Fluid Mech.}  \bvol{798},  \pg{R2}.

\bibitem[Halko {\em et~al.\/}(2011)Halko, Martinsson \& Tropp]{Halko:SIAM11}
{\sc \au{Halko, N.}, \au{Martinsson, P.-G.} \& \au{Tropp, J.~A.}} \yr{2011}
  \at{Finding structure with randomness: Probabilistic algorithms for
  constructing approximate matrix decompositions}.  \jt{SIAM review}
  \bvol{53}~(2),  \pg{217--288}.

\bibitem[Harper \& Maki(1964)]{Harper:64}
{\sc \au{Harper, C.~W.} \& \au{Maki, R.~L.}} \yr{1964}  \bt{A review of the
  stall characteristics of swept wings}. {\em Tech. Rep.\/} NASA/TN D-2373.
  \org{NASA, Washington, DC}.

\bibitem[He \& Timme(2021)]{HeTimme:JFM21}
{\sc \au{He, W.} \& \au{Timme, S.}} \yr{2021}  \at{Triglobal infinite-wing
  shock-buffet study}.  \jt{J. Fluid Mech.}  \bvol{925},  \pg{A27}.

\bibitem[Hill(1992)]{Hill:AIAA92}
{\sc \au{Hill, D.}} \yr{1992}  \bt{A theoretical approach for analyzing the
  restabilization of wakes}.  \publ{AIAA Paper 1992-67}.

\bibitem[House {\em et~al.\/}(2022)House, Skene, Ribeiro, Yeh \&
  Taira]{House:AIAA22}
{\sc \au{House, D.~C.}, \au{Skene, C.~S.}, \au{Ribeiro, J. H.~M.}, \au{Yeh,
  C.-A.} \& \au{Taira, K.}} \yr{2022}  \bt{Sketch-based resolvent analysis}.
  \publ{AIAA Paper 2022-3335}.

\bibitem[Houtman {\em et~al.\/}(2022)Houtman, Timme \& Sharma]{Houtman:AIAA22}
{\sc \au{Houtman, J.}, \au{Timme, S.} \& \au{Sharma, A.}} \yr{2022}
  \bt{Resolvent analysis of large aircraft wings in edge-of-the-envelope
  transonic flow}.  \publ{AIAA Paper 2022-1329}.

\bibitem[Jovanovi{\'c}(2004)]{Jovanovic:04}
{\sc \au{Jovanovi{\'c}, M.~R.}} \yr{2004} {\em Modeling, analysis, and control
  of spatially distributed systems\/}.  \publ{University of California at Santa
  Barbara, Dept. of Mechanical Engineering}.

\bibitem[Jovanovi{\'c} \& Bamieh(2005)]{Jovanovic:JFM05}
{\sc \au{Jovanovi{\'c}, M.~R.} \& \au{Bamieh, B.}} \yr{2005}  \at{Componentwise
  energy amplification in channel flows}.  \jt{J. Fluid Mech.}  \bvol{534},
  \pg{145--183}.

\bibitem[Khalighi {\em et~al.\/}(2011)Khalighi, Ham, Nichols, Lele \&
  Moin]{Khalighi:AIAA11}
{\sc \au{Khalighi, Y.}, \au{Ham, F.}, \au{Nichols, J.}, \au{Lele, S.~K.} \&
  \au{Moin, P.}} \yr{2011}  \bt{Unstructured large eddy simulation for
  prediction of noise issued from turbulent jets in various configurations}.
  \publ{AIAA Paper 2011--2886}.

\bibitem[Lentink {\em et~al.\/}(2007)Lentink, M{\"u}ller, Stamhuis, {De Kat},
  {Van Gestel}, Veldhuis, Henningsson, Hedenstr{\"o}m, Videler \& {Van
  Leeuwen}]{Lentink:Nature07}
{\sc \au{Lentink, D.}, \au{M{\"u}ller, U.~K.}, \au{Stamhuis, E.~J.}, \au{{De
  Kat}, R.}, \au{{Van Gestel}, W.}, \au{Veldhuis, L. L.~M.}, \au{Henningsson,
  P.}, \au{Hedenstr{\"o}m, A.}, \au{Videler, J.~J.} \& \au{{Van Leeuwen},
  J.~L.}} \yr{2007}  \at{How swifts control their glide performance with
  morphing wings}.  \jt{Nature}  \bvol{446}~(7139),  \pg{1082--1085}.

\bibitem[Liu {\em et~al.\/}(2021)Liu, Sun, Yeh, Ukeiley, Cattafesta \&
  Taira]{Liu:JFM21}
{\sc \au{Liu, Q.}, \au{Sun, Y.}, \au{Yeh, C.-A.}, \au{Ukeiley, L.~S.},
  \au{Cattafesta, L.~N.} \& \au{Taira, K.}} \yr{2021}  \at{Unsteady control of
  supersonic turbulent cavity flow based on resolvent analysis}.  \jt{J. Fluid
  Mech.}  \bvol{925},  \pg{A5}.

\bibitem[Martini {\em et~al.\/}(2021)Martini, Rodr{\'\i}guez, Towne \&
  Cavalieri]{Martini:JFM21}
{\sc \au{Martini, E.}, \au{Rodr{\'\i}guez, D.}, \au{Towne, A.} \&
  \au{Cavalieri, A. V.~G.}} \yr{2021}  \at{Efficient computation of global
  resolvent modes}.  \jt{J. Fluid Mech.}  \bvol{919},  \pg{A3}.

\bibitem[Masini {\em et~al.\/}(2020)Masini, Timme \& Peace]{Masini:JFM20}
{\sc \au{Masini, L.}, \au{Timme, S.} \& \au{Peace, A.~J.}} \yr{2020}
  \at{Analysis of a civil aircraft wing transonic shock buffet experiment}.
  \jt{J. Fluid Mech.}  \bvol{884},  \pg{A1}.

\bibitem[Mc{K}eon \& Sharma(2010)]{McKeon:JFM10}
{\sc \au{Mc{K}eon, B.~J.} \& \au{Sharma, A.~S.}} \yr{2010}  \at{A
  critical-layer framework for turbulent pipe flow}.  \jt{J. Fluid Mech.}
  \bvol{658},  \pg{336--382}.

\bibitem[Moarref {\em et~al.\/}(2013)Moarref, Sharma, Tropp \&
  McKeon]{Moarref:JFM13}
{\sc \au{Moarref, R.}, \au{Sharma, A.~S.}, \au{Tropp, J.~A.} \& \au{McKeon,
  B.~J.}} \yr{2013}  \at{Model-based scaling of the streamwise energy density
  in high-reynolds-number turbulent channels}.  \jt{J. Fluid Mech.}
  \bvol{734},  \pg{275--316}.

\bibitem[Monokrousos {\em et~al.\/}(2010)Monokrousos, {\AA}kervik, Brandt \&
  Henningson]{Monokrousos:JFM10}
{\sc \au{Monokrousos, A.}, \au{{\AA}kervik, E.}, \au{Brandt, L.} \&
  \au{Henningson, D.~S.}} \yr{2010}  \at{Global three-dimensional optimal
  disturbances in the blasius boundary-layer flow using time-steppers}.  \jt{J.
  Fluid Mech.}  \bvol{650},  \pg{181--214}.

\bibitem[Paladini {\em et~al.\/}(2019)Paladini, Beneddine, Dandois, Sipp \&
  Robinet]{Paladini:PRF19}
{\sc \au{Paladini, E.}, \au{Beneddine, S.}, \au{Dandois, J.}, \au{Sipp, D.} \&
  \au{Robinet, J.-C.}} \yr{2019}  \at{Transonic buffet instability: From
  two-dimensional airfoils to three-dimensional swept wings}.  \jt{Phys. Rev.
  Fluids}  \bvol{4}~(10),  \pg{103906}.

\bibitem[Plante {\em et~al.\/}(2021)Plante, Dandois, Beneddine, Laurendeau \&
  Sipp]{Plante:JFM21}
{\sc \au{Plante, F.}, \au{Dandois, J.}, \au{Beneddine, S.}, \au{Laurendeau,
  {\'E}.} \& \au{Sipp, D.}} \yr{2021}  \at{Link between subsonic stall and
  transonic buffet on swept and unswept wings: from global stability analysis
  to nonlinear dynamics}.  \jt{J. Fluid Mech.}  \bvol{908},  \pg{A16}.

\bibitem[Qadri \& Schmid(2017)]{Qadri:PRF17}
{\sc \au{Qadri, U.~A.} \& \au{Schmid, P.~J.}} \yr{2017}  \at{Frequency
  selection mechanisms in the flow of a laminar boundary layer over a shallow
  cavity}.  \jt{Phys. Rev. Fluids}  \bvol{2},  \pg{043902}.

\bibitem[Ribeiro {\em et~al.\/}(2020)Ribeiro, Yeh \& Taira]{Ribeiro:PRF20}
{\sc \au{Ribeiro, J. H.~M.}, \au{Yeh, C.-A.} \& \au{Taira, K.}} \yr{2020}
  \at{Randomized resolvent analysis}.  \jt{Phys. Rev. Fluids}  \bvol{5}~(3),
  \pg{033902}.

\bibitem[Ribeiro {\em et~al.\/}(2022{\natexlab{{\em a\/}}})Ribeiro, Yeh, Zhang
  \& Taira]{Ribeiro:AIAA22}
{\sc \au{Ribeiro, J. H.~M.}, \au{Yeh, C.-A.}, \au{Zhang, K.} \& \au{Taira, K.}}
  \yr{2022{\natexlab{{\em a\/}}}}  \bt{From biglobal to triglobal resolvent
  analysis: laminar separated flows over swept wings}.  \publ{AIAA Paper
  2022-2428}.

\bibitem[Ribeiro {\em et~al.\/}(2022{\natexlab{{\em b\/}}})Ribeiro, Yeh, Zhang
  \& Taira]{Ribeiro:JFM22}
{\sc \au{Ribeiro, J. H.~M.}, \au{Yeh, C.-A.}, \au{Zhang, K.} \& \au{Taira, K.}}
  \yr{2022{\natexlab{{\em b\/}}}}  \at{Wing sweep effects on laminar separated
  flows}.  \jt{J. Fluid Mech.}  \bvol{950},  \pg{A23}.

\bibitem[Ricciardi {\em et~al.\/}(2022)Ricciardi, Wolf \&
  Taira]{Ricciardi:JFM22}
{\sc \au{Ricciardi, T.~R.}, \au{Wolf, W.~R.} \& \au{Taira, K.}} \yr{2022}
  \at{Transition, intermittency and phase interference effects in airfoil
  secondary tones and acoustic feedback loop}.  \jt{J. Fluid Mech.}
  \bvol{937}.

\bibitem[Schmid \& Brandt(2014)]{Schmid:AMR14}
{\sc \au{Schmid, P.~J.} \& \au{Brandt, L.}} \yr{2014}  \at{Analysis of fluid
  systems: Stability, receptivity, sensitivity}.  \jt{Applied Mechanics
  Reviews}  \bvol{66}~(2), 024803.

\bibitem[Schmidt {\em et~al.\/}(2018)Schmidt, Towne, Rigas, Colonius \&
  Br{\`e}s]{Schmidt:JFM18}
{\sc \au{Schmidt, O.~T.}, \au{Towne, A.}, \au{Rigas, G.}, \au{Colonius, T.} \&
  \au{Br{\`e}s, G.~A.}} \yr{2018}  \at{Spectral analysis of jet turbulence}.
  \jt{J. Fluid Mech.}  \bvol{855},  \pg{953--982}.

\bibitem[Skene {\em et~al.\/}(2022{\natexlab{{\em a\/}}})Skene, Ribeiro \&
  Taira]{SkeneRibeiro:tools}
{\sc \au{Skene, C.~S.}, \au{Ribeiro, J. H.~M.} \& \au{Taira, K.}}
  \yr{2022{\natexlab{{\em a\/}}}} csskene/linear-analysis-tools: Initial
  release. https://doi.org/10.5281/zenodo.6550726.

\bibitem[Skene \& Schmid(2019)]{Skene:JFM19}
{\sc \au{Skene, C.~S.} \& \au{Schmid, P.~J.}} \yr{2019}  \at{Adjoint-based
  parametric sensitivity analysis for swirling {M}-flames}.  \jt{J. Fluid
  Mech.}  \bvol{859},  \pg{516--542}.

\bibitem[Skene {\em et~al.\/}(2022{\natexlab{{\em b\/}}})Skene, Yeh, Schmid \&
  Taira]{Skene:JFM22}
{\sc \au{Skene, C.~S.}, \au{Yeh, C.-A.}, \au{Schmid, P.~J.} \& \au{Taira, K.}}
  \yr{2022{\natexlab{{\em b\/}}}}  \at{Sparsifying the resolvent forcing mode
  via gradient-based optimisation}.  \jt{J. Fluid Mech.}  \bvol{944},
  \pg{A52}.

\bibitem[Stewart(2002)]{Stewart:SIAM02}
{\sc \au{Stewart, G.~W.}} \yr{2002}  \at{A {K}rylov--{S}chur algorithm for
  large eigenproblems}.  \jt{SIAM J. Mat. Anal. App.}  \bvol{23}~(3),
  \pg{601--614}.

\bibitem[Strykowski \& Sreenivasan(1990)]{Strykowski:90}
{\sc \au{Strykowski, P.~J.} \& \au{Sreenivasan, K.~R.}} \yr{1990}  \at{On the
  formation and suppression of vortex ‘shedding’at low reynolds numbers}.
  \jt{J. Fluid Mech.}  \bvol{218},  \pg{71--107}.

\bibitem[Sun {\em et~al.\/}(2017)Sun, Taira, III \& Ukeiley]{Sun:JFM17}
{\sc \au{Sun, Y.}, \au{Taira, K.}, \au{III, L. N.~Cattafesta} \& \au{Ukeiley,
  L.~S.}} \yr{2017}  \at{Biglobal instabilities of compressible open-cavity
  flows}.  \jt{J. Fluid Mech.}  \bvol{826},  \pg{270--301}.

\bibitem[Taira {\em et~al.\/}(2017)Taira, Brunton, Dawson, Rowley, Colonius,
  McKeon, Schmidt, Gordeyev, Theofilis \& Ukeiley]{Taira:AIAAJ17}
{\sc \au{Taira, K.}, \au{Brunton, S.~L.}, \au{Dawson, S. T.~M.}, \au{Rowley,
  C.~W.}, \au{Colonius, T.}, \au{McKeon, B.~J.}, \au{Schmidt, O.~T.},
  \au{Gordeyev, S.}, \au{Theofilis, V.} \& \au{Ukeiley, L.~S.}} \yr{2017}
  \at{Modal analysis of fluid flows: An overview}.  \jt{AIAA J.}
  \bvol{55}~(12),  \pg{4013--4041}.

\bibitem[Taira \& Colonius(2009)]{Taira:JFM09}
{\sc \au{Taira, K.} \& \au{Colonius, T.}} \yr{2009}  \at{Three-dimensional
  flows around low-aspect-ratio flat-plate wings at low {R}eynolds numbers}.
  \jt{J. Fluid Mech.}  \bvol{623},  \pg{187--207}.

\bibitem[Taira {\em et~al.\/}(2020)Taira, Hemati, Brunton, Sun, Duraisamy,
  Bagheri, Dawson \& Yeh]{Taira:AIAAJ20}
{\sc \au{Taira, K.}, \au{Hemati, M.~S.}, \au{Brunton, S.~L.}, \au{Sun, Y.},
  \au{Duraisamy, K.}, \au{Bagheri, S.}, \au{Dawson, S. T.~M.} \& \au{Yeh,
  C.-A.}} \yr{2020}  \at{Modal analysis of fluid flows: Applications and
  outlook}.  \jt{AIAA J.}  \bvol{58}~(3),  \pg{998--1022}.

\bibitem[Thomareis \& Papadakis(2018)]{Thomareis:PRF18}
{\sc \au{Thomareis, N.} \& \au{Papadakis, G.}} \yr{2018}  \at{Resolvent
  analysis of separated and attached flows around an airfoil at transitional
  reynolds number}.  \jt{Phys. Rev. Fluids}  \bvol{3}~(7),  \pg{073901}.

\bibitem[Timme(2020)]{Timme:JFM20}
{\sc \au{Timme, S.}} \yr{2020}  \at{Global instability of wing shock-buffet
  onset}.  \jt{J. Fluid Mech.}  \bvol{885},  \pg{A37}.

\bibitem[Torres \& Mueller(2004)]{Torres:AIAAJ04}
{\sc \au{Torres, G.~E.} \& \au{Mueller, T.~J.}} \yr{2004}  \at{Low-aspect-ratio
  aerodynamics at low {R}eynolds numbers}.  \jt{AIAA J.}  \bvol{42}~(5),
  \pg{865--873}.

\bibitem[Trefethen {\em et~al.\/}(1993)Trefethen, Trefethen, Reddy \&
  Driscoll]{Trefethen:93}
{\sc \au{Trefethen, L.~N.}, \au{Trefethen, A.~E.}, \au{Reddy, S.~C.} \&
  \au{Driscoll, T.~A.}} \yr{1993}  \at{Hydrodynamic stability without
  eigenvalues}.  \jt{Science}  \bvol{261}~(5121),  \pg{578--584}.

\bibitem[Videler {\em et~al.\/}(2004)Videler, Stamhuis \&
  Povel]{Videler:Science04}
{\sc \au{Videler, J.~J.}, \au{Stamhuis, E.~J.} \& \au{Povel, G. D.~E.}}
  \yr{2004}  \at{Leading-edge vortex lifts swifts}.  \jt{Science}
  \bvol{306}~(5703),  \pg{1960--1962}.

\bibitem[Wygnanski {\em et~al.\/}(2011)Wygnanski, Tewes, Kurz, Taubert \&
  Chen]{Wygnanski:JFM11}
{\sc \au{Wygnanski, I.}, \au{Tewes, P.}, \au{Kurz, H.}, \au{Taubert, L.} \&
  \au{Chen, C.}} \yr{2011}  \at{The application of boundary layer independence
  principle to three-dimensional turbulent mixing layers}.  \jt{J. Fluid Mech.}
   \bvol{675},  \pg{336--346}.

\bibitem[Yeh {\em et~al.\/}(2020)Yeh, Benton, Taira \& Garmann]{Yeh:PRF20}
{\sc \au{Yeh, C.-A.}, \au{Benton, S.~I.}, \au{Taira, K.} \& \au{Garmann,
  D.~J.}} \yr{2020}  \at{Resolvent analysis of an airfoil laminar separation
  bubble at {Re} = 500 000}.  \jt{Phys. Rev. Fluids}  \bvol{5}~(8),
  \pg{083906}.

\bibitem[Yeh \& Taira(2019)]{Yeh:JFM19}
{\sc \au{Yeh, C.-A.} \& \au{Taira, K.}} \yr{2019}  \at{Resolvent-analysis-based
  design of airfoil separation control}.  \jt{J. Fluid Mech.}  \bvol{867},
  \pg{572--610}.

\bibitem[Yen \& Hsu(2007)]{Yen:AIAAJ07}
{\sc \au{Yen, S.-C.} \& \au{Hsu, C.~M.}} \yr{2007}  \at{Flow patterns and wake
  structure of a swept-back wing}.  \jt{AIAA J.}  \bvol{45}~(1),
  \pg{228--236}.

\bibitem[Yen \& Huang(2009)]{Yen:JFE09}
{\sc \au{Yen, S.-C.} \& \au{Huang, L.-C.}} \yr{2009}  \at{Flow patterns and
  aerodynamic performance of unswept and swept-back wings}.  \jt{J. Fluids
  Eng.}  \bvol{131}~(11).

\bibitem[Yilmaz \& Rockwell(2012)]{Yilmaz:JFM12}
{\sc \au{Yilmaz, T.~O.} \& \au{Rockwell, D.}} \yr{2012}  \at{Flow structure on
  finite-span wings due to pitch-up motion}.  \jt{J. Fluid Mech.}  \bvol{691},
  \pg{518--545}.

\bibitem[Zhang {\em et~al.\/}(2020{\natexlab{{\em a\/}}})Zhang, Hayostek,
  Amitay, Burstev, Theofilis \& Taira]{Zhang:JFM20b}
{\sc \au{Zhang, K.}, \au{Hayostek, S.}, \au{Amitay, M.}, \au{Burstev, A.},
  \au{Theofilis, V.} \& \au{Taira, K.}} \yr{2020{\natexlab{{\em a\/}}}}
  \at{Laminar separated flows over finite-aspect-ratio swept wings}.  \jt{J.
  Fluid Mech.}  \bvol{905},  \pg{R1}.

\bibitem[Zhang {\em et~al.\/}(2020{\natexlab{{\em b\/}}})Zhang, Hayostek,
  Amitay, He, Theofilis \& Taira]{Zhang:JFM20}
{\sc \au{Zhang, K.}, \au{Hayostek, S.}, \au{Amitay, M.}, \au{He, W.},
  \au{Theofilis, V.} \& \au{Taira, K.}} \yr{2020{\natexlab{{\em b\/}}}}  \at{On
  the formation of three-dimensional separated flows over wings under tip
  effects}.  \jt{J. Fluid Mech.}  \bvol{895},  \pg{A9}.

\bibitem[Zhang \& Taira(2022)]{Zhang:PRF22}
{\sc \au{Zhang, K.} \& \au{Taira, K.}} \yr{2022}  \at{Laminar vortex dynamics
  around forward-swept wings}.  \jt{Phys. Rev. Fluids}  \bvol{7}~(2),
  \pg{024704}.

\end{thebibliography}
	\bibliographystyle{jfm}
	
\end{document}